\documentstyle[12pt,epsf,cite,subeqnarray]{article}
\newcommand{\gev}{\,{\rm GeV}}
\newcommand{\tev}{\,{\rm TeV}}
\newcommand{\lsim}{
  \raisebox{0.2em}{$<$} \hspace{-0.75em} \raisebox{-0.2em}{$\sim$} }

\makeatletter

\renewcommand{\appendix}{%
\par
\setcounter{section}{0}
\setcounter{subsection}{0}
\@addtoreset{equation}{section}
\@addtoreset{equation}{subsection}
\renewcommand{\thesection}{Appendix~\Alph{section}}
\renewcommand{\thesubsection}{\Alph{section}.\arabic{subsection}}
\renewcommand{\thesubsubsection}%
{\Alph{section}.\arabic{subsection}.\arabic{subsubsection}}
\renewcommand{\theequation}%
{\Alph{section}.\arabic{subsection}.\arabic{equation}}
}

\newcommand{\arraystrut}{%
\protect\rule[-1.4ex]{0em}{4.2ex}\protect\rule{\arraycolsep}{0ex}%
}
\newcommand{\ie}{{\it i.e.\/}}
\newcommand{\ol}[1]{\overline{#1}}
\newcommand{\wt}[1]{\widetilde{#1}}
\newcommand{\vev}[1]{\left\langle #1 \right\rangle}
\newcommand{\hc}{{\rm H.\,c.}\,}
\newcommand{\PL}{{\bf L}}
\newcommand{\PR}{{\bf R}}
\newcommand{\Dt}{(4\pi)^2\Lambda\frac{d}{d\Lambda}}

\makeatother
\title{ \vspace*{-10mm}
\begin{tabular}{ll}
\hspace*{8cm} & {\normalsize KEK-TH-583} \vspace{-3mm} \\
\hspace*{8cm} & {\normalsize KEK Preprint 98-122} \vspace{-3mm} \\
\hspace*{8cm} & {\normalsize August 1998}
\end{tabular} \vspace{2mm} \\
\bf Effect of an $RRRR$ dimension 5 operator on the proton decay 
in the minimal SU(5) SUGRA GUT model} 
\author{Toru Goto and Takeshi Nihei\\
 \mbox{} \vspace{3mm}\\
 \normalsize \em  Theory Group, KEK, Tsukuba, Ibaraki 305-0801, Japan }
\date{ }
\begin{document}
\baselineskip 18pt
\renewcommand{\thefootnote}{\fnsymbol{footnote}}
\begin{titlepage}
\maketitle
\thispagestyle{empty}
\begin{abstract}
\normalsize
\baselineskip 18pt
We reanalyze the proton decay in the minimal 
SU(5) SUGRA GUT model. 
Unlike previous analyses, we take into account 
a Higgsino dressing diagram of 
dimension 5 operator with right-handed 
matter fields ($RRRR$ operator). 
It is shown that this diagram gives a dominant
contribution for 
$p\rightarrow K^+\overline{\nu}_\tau$ over that from 
$LLLL$ operator, 
and decay rate of this mode 
can be comparable with that of 
$p\rightarrow K^+\overline{\nu}_\mu$ which is dominated by the
$LLLL$ contribution. 
It is found that 
we cannot reduce both the decay rate of 
$p\rightarrow K^+\overline{\nu}_\tau$ and 
that of $p\rightarrow K^+\overline{\nu}_\mu$ simultaneously
by adjusting relative phases between Yukawa couplings
at colored Higgs interactions. 
Constraints on the colored Higgs mass $M_C$ and 
a typical squark and slepton mass $m_{\tilde{f}}$ 
from Super-Kamiokande limit 
become considerably stronger due to the Higgsino dressing diagram of 
the $RRRR$ operator: 
$M_C$ $>$ $6.5 \times 10^{16} \gev$ for $m_{\tilde{f}}$ $<$ $1 \tev$, 
and $m_{\tilde{f}}$ $>$ $2.5 \tev$ for $M_C$ $<$ $2.5 \times 10^{16} \gev$. 
\end{abstract}
\end{titlepage}
\newpage
%
%
\section{Introduction}
%
\hspace*{0.5cm}
%
The gauge coupling unification around $M_X \sim 2 \times 10^{16} \gev$ 
\cite{Gauge_Coupling_Unification}
strongly suggests the 
supersymmetric grand unified theory (SUSY GUT) \cite{SUSY_GUT}.
In this model, the gauge hierarchy problem is naturally
solved by supersymmetry. 
Also, 
this model makes successful predictions for 
the charge quantization and the bottom-tau mass ratio. 
Proton decay is one of the direct consequences of grand unification. 
The main decay mode $ p \rightarrow K^{+}\overline{\nu} $ \cite{DRW,ENR} 
in the minimal SU(5) supergravity (SUGRA) GUT model \cite{SUGRA} 
has been searched for with the underground experiments \cite{Kam,IMB},
and the previous results have already imposed severe
constraints on this model. 
Recently new results of the proton decay search at 
Super-Kamiokande have been reported \cite{superK}. The bound on the
partial lifetime of the $K^+\overline{\nu}$ mode is
 $\tau(p\rightarrow K^+ \overline{\nu})$
$>$ $5.5 \times 10^{32}$ years (90\,\% C.L.),
where three neutrinos are not distinguished.

There are a number of detailed analyses on 
the nucleon decay in the minimal SU(5) SUGRA GUT 
model \cite{dim5_op,DRW,ENR,NCA,MATS+HMY,HMTY,GNA}.
In the previous analyses, it is believed that 
contribution from dimension 5 operator with left-handed 
matter fields ($LLLL$ operator) is dominant 
for $ p \rightarrow K^{+} \overline{\nu} $ \cite{ENR}. 
In particular 
a Higgsino dressing diagram of $RRRR$ operator has been ignored 
in these analyses. 
It has been concluded that 
the main decay mode is 
$ p \rightarrow K^{+} \overline{\nu}_\mu $ \cite{DRW}, 
and the decay rate of this mode 
can be suppressed sufficiently 
by adjusting relative
phases between Yukawa couplings
at colored Higgs interactions \cite{NCA}. 
Recently it has been pointed out that 
the Higgsino dressing diagram of the $RRRR$ operator 
gives a significant contribution 
to $p\rightarrow K^+\overline{\nu}_\tau$ 
in a large $\tan \beta$ region
in the context of a SUSY SO(10) GUT model \cite{RRRR}.

In this paper, we reanalyze the proton decay 
including the $RRRR$ operator 
in the minimal SU(5) SUGRA GUT model.
We calculate all the dressing diagrams \cite{NCA} (exchanging
the charginos, the neutralinos and the gluino) of 
the $LLLL$ and $RRRR$ operators, 
taking account of various mixing effects among the SUSY particles, 
such as flavor mixing of quarks and squarks, left-right mixing of 
squarks and sleptons, and gaugino-Higgsino mixing of charginos 
and neutralinos. 
For this purpose we diagonalize mass matrices numerically to
obtain the mixing factors at `ino' vertices 
and the dimension 5 couplings. 
We examine the effect of the relative
phases between the Yukawa couplings
at the colored Higgs interactions. 
We show that the Higgsino dressing diagram of 
the $RRRR$ operator gives a dominant
contribution for 
$p\rightarrow K^+\overline{\nu}_\tau$,
and the decay rate of this mode 
can be comparable with that of 
$p\rightarrow K^+\overline{\nu}_\mu$ which is dominated by the
$LLLL$ contribution. 
We find that 
we cannot reduce both the decay rate of 
$p\rightarrow K^+\overline{\nu}_\tau$ and 
that of $p\rightarrow K^+\overline{\nu}_\mu$ simultaneously
by adjusting the relative phases. 
We obtain constraints on the colored Higgs mass
and the typical mass scale of squarks and sleptons 
under the updated Super-Kamiokande bound, 
and find that these constraints are much stronger than 
that derived from the analysis neglecting the $RRRR$ effect. 

This paper is organized as follows. In Section 2, we descibe 
the dimension 5 operators in the minimal SU(5) SUGRA GUT 
and briefly sketch our scheme to calculate the proton decay rates. 
We give a qualitative discussion on the $RRRR$ contribution in Section 3.
We present results of our numerical calculation and discuss 
constraints on this model in Section 4. 
Formulas used in the calculation of the nucleon decay rates are 
summarized in Appendix A. 

%
%
%
\section{Dimension 5 operators in the minimal SU(5) SUGRA GUT}
%
\hspace*{0.5cm}
%
%
%
Nucleon decay in the minimal SU(5) SUGRA GUT model is 
dominantly caused by dimension 5 operators \cite{dim5_op}, which 
are generated by the exchange of the colored Higgs multiplet. 
The dimension 5 operators relevant to the
nucleon decay are described by the following 
superpotential: 
\begin{eqnarray}
W_5 & = & -\frac{1}{M_C} \left\{ \frac{1}{2}C_{5L}^{ijkl}
Q_k Q_l Q_i L_j + C_{5R}^{ijkl}E^c_k U^c_l U^c_i D^c_j  \right\}.
\label{eqn:dim5_op}
\end{eqnarray}
Here $Q$, $U^c$ and $E^c$ are chiral superfields which contain 
a left-handed quark doublet, 
a charge conjugation of a right-handed up-type quark, 
and a charge conjugation of a right-handed charged lepton, 
respectively, and are embedded in the 10 
representation of SU(5). The chiral superfields $L$ and $D^c$ contain 
a left-handed lepton doublet and 
a charge conjugation of a right-handed down-type quark,
respectively, and are embedded in the $\overline{5}$ 
representation. 
A mass of the colored Higgs superfields is denoted by $M_C$. 
The indices $ i,j,k,l = 1,2,3$ are generation labels. 
The first term in Eq.~(\ref{eqn:dim5_op}) 
represents $LLLL$ operator \cite{ENR} which 
contains only left-handed
SU(2) doublets. The second term
in Eq.~(\ref{eqn:dim5_op}) represents $RRRR$ operator which 
 contains only right-handed SU(2) singlets.
%
The coefficients $C_{5L}$ and $C_{5R}$ 
in Eq.~(\ref{eqn:dim5_op}) are determined by 
Yukawa coupling matrices \cite{NCA}. 
Approximately these are written as 
\begin{eqnarray}
\left. C_{5L}^{ijkl} \right|_X 
&\approx& (Y_D)_{ij} (V^{\bf T} P Y_U V)_{kl}, \nonumber \\
\left. C_{5R}^{ijkl} \right|_X &\approx& (P^*V^*Y_D)_{ij}(V^{\bf T} Y_U)_{kl},
\label{eqn:C5L&C5R}
\end{eqnarray}
where $Y_U$ and $Y_D$ 
are diagonalized
Yukawa coupling matrices for $10 \cdot 10 \cdot 5_H$ and 
$10 \cdot \overline{5} \cdot \overline{5}_H$ interactions,
respectively. 
More precise expressions for  $C_{5L}$ and $C_{5R}$ are given 
in Appendix A. 
The unitary matrix $V$ is the CKM matrix at the GUT scale.
The matrix $P$ $=$ ${\rm diag}(P_1,P_2,P_3)$ is a 
diagonal unimodular phase matrix with $|P_i|=1$ and $ {\rm det}P=1$. 
We parametrize $P$ as 
\begin{eqnarray}
P_1/P_3 = e^{i\phi_{13}}, \hspace{3mm}
P_2/P_3 = e^{i\phi_{23}}.
\label{eqn:Phase_Matrix}
\end{eqnarray}
The parameters $\phi_{13}$ and $\phi_{23}$ are 
relative phases between the Yukawa couplings 
at the colored Higgs interactions, and 
cannot be removed by field redefinitions \cite{Phase_Matrix}. 
The expressions for $C_{5L}$ and $C_{5R}$ in Eq.~(\ref{eqn:C5L&C5R}) 
are written in the flavor basis 
where the Yukawa coupling matrix for 
the $10 \cdot \overline{5} \cdot \overline{5}_H$ interaction 
is diagonalized at the GUT scale. 
Numerical values of $Y_U$, $Y_D$ and $V$ at the GUT scale are
calculated from the quark masses and the CKM matrix 
at the electroweak scale 
using renormalization group equations (RGEs).

In the minimal SU(5) SUGRA GUT, 
soft SUSY breaking parameters at the Planck scale 
are described by $m_0$, $M_{gX}$ and $A_X$
which denote universal scalar mass, universal gaugino mass, 
and universal coefficient of the trilinear scalar couplings, 
respectively. 
Low energy values of the soft breaking parameters are 
determined by solving the one-loop RGEs \cite{BKS}. 
The electroweak symmetry is broken radiatively \cite{Radiative_Breaking}
due to the effect of a large Yukawa coupling of the top quark,
and we require that the correct vacuum expectation values of 
the Higgs fields at the electroweak scale are reproduced. 
We ignore RGE running effects between the Planck scale and the GUT
scale for simplicity. 
In this approximation the phase matrix $P$ decouples from the 
RGEs of the soft SUSY breaking parameters.
Thus we have all the values of the parameters at the electroweak
scale. 
The masses and the mixings are obtained by diagonalizing
the mass matrices numerically. 
We evaluate hadronic matrix elements using the chiral Lagrangian 
method \cite{Chiral_Lagrangian}.
The parameters $\alpha_p$ and $\beta_p$ defined by
$\langle 0| \epsilon_{\hat{a}\hat{b}\hat{c}} 
(d_R^{\hat{a}} u_R^{\hat{b}}) u_L^{\hat{c}} |p 
\rangle$ $=$ $\alpha_p N_L$ and 
$\langle 0| \epsilon_{\hat{a}\hat{b}\hat{c}} 
(d_L^{\hat{a}} u_L^{\hat{b}}) u_L^{\hat{c}} |p 
\rangle$ $=$ $\beta_p N_L$
($N_L$ is a left-handed proton's wave function) 
are evaluated as $0.003 \, {\rm GeV}^3$ $\leq$ $\beta_p$
$\leq$ $0.03 \, {\rm GeV}^3$ and $\alpha_p$ $=$ $-\beta_p$ 
by various methods \cite{beta_p}. 
We use the smallest value 
$\beta_p$ $=$ $-\alpha_p$ $=$ $0.003 \, {\rm GeV}^3$ 
in our analysis to obtain conservative bounds. 
For the details of the methods of our analysis, 
see Ref.~\cite{GNA,RRRR}. 
Formulas for relevant interactions and the nucleon decay rates 
are given in Appendix A. 

%
%
%
\section{$RRRR$ contribution to the proton decay}
\begin{figure}[p]
\hspace*{5mm}
\leavevmode\epsffile{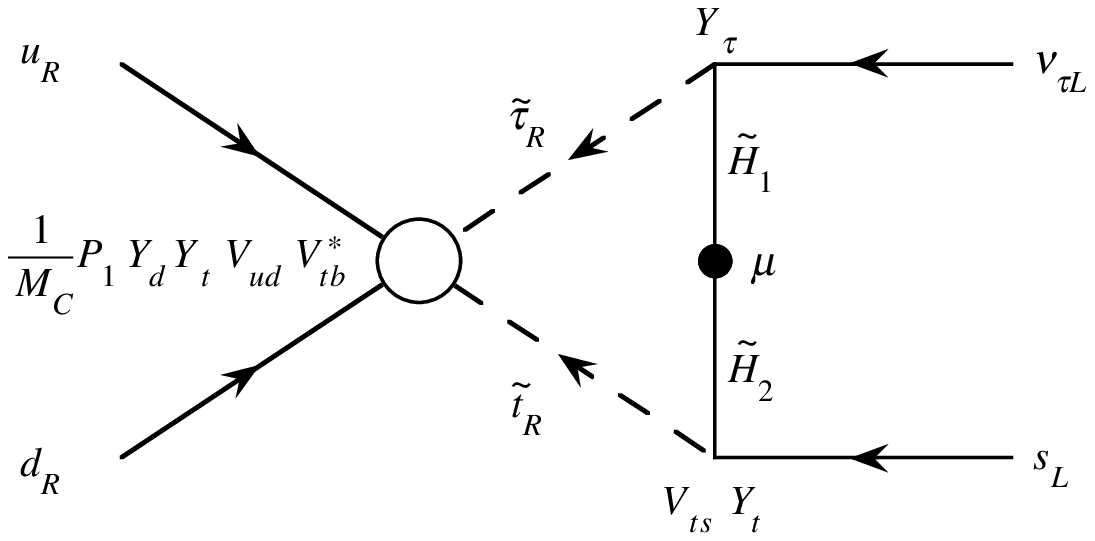}
\caption[figI]{Higgsino dressing diagram which gives a dominant
contribution to the $p\rightarrow K^+ \overline{\nu}_\tau$ mode. 
The circle represents
the $RRRR$ dimension 5 operator. We also have a similar diagram 
for $(u_R s_R)(d_L \nu_{\tau L})$.}
\label{fig:diagram}
\end{figure}

%
\hspace*{0.5cm}  
%
%
%
The dimension 5 operators consist of two fermions and two bosons.
Eliminating the two scalar bosons by the gaugino or Higgsino
exchange (dressing), we obtain the four-fermion interactions
which cause the nucleon decay \cite{ENR,NCA}.
In the one-loop calculations of the dressing diagrams,
we include all the dressing diagrams exchanging
the charginos, the neutralinos and the gluino of the
$LLLL$ and $RRRR$ dimension 5 operators. 
In addition to the contributions from the dimension 5 operators, 
we include the contributions from 
dimension 6 operators mediated by the heavy gauge boson
and the colored Higgs boson. 
Though the effects of the dimension 6 operators ($\sim$ $1/M_X^2$)
are negligibly small for
$p\rightarrow K^+\overline{\nu}$, 
these could be important for other decay modes such as 
$p\rightarrow \pi^0 e^+$. 
%
The major contribution of the $LLLL$ operator comes from
an ordinary diagram with wino dressing. 
The major contribution of the $RRRR$ operator arises from
a Higgsino dressing diagram depicted in Fig.~\ref{fig:diagram}. 
The circle in this figure represents 
the complex conjugation of $C_{5R}^{ijkl}$ in Eq.~(\ref{eqn:C5L&C5R}) 
with $i$ $=$ $j$ $=1$ and $k$ $=$ $l$ $=3$. 
This diagram contains the Yukawa couplings
of the top quark and the tau lepton. 
Importance of this diagram
has already been pointed out in Ref.~\cite{RRRR} in the context of 
a SUSY SO(10) GUT model. 
This diagram has been ignored 
in previous analyses in the minimal SU(5) SUSY 
GUT \cite{dim5_op,DRW,ENR,NCA,MATS+HMY,HMTY,GNA},
though the contributions from gaugino dressing
of the $RRRR$ operator were included
in Ref.~\cite{NCA}. 
We show that this diagram indeed 
gives a significant contribution in the case of the 
minimal SU(5) SUGRA GUT model also.

Before we proceed to present results of our numerical calculations, 
we give a rough estimation for the decay amplitudes 
for a qualitative understanding of the results. 
In the actual calculations, however, we make full numerical analyses 
including contributions from all the dressing diagrams as well as 
those from dimension six operators.
We also take account of 
various effects such as mixings between the SUSY particles. 
Besides the soft breaking parameter dependence arising from the loop
calculations, relative magnitudes 
between various contributions can be roughly 
understood by the form of the dimension 5 operator 
in Eq.~(\ref{eqn:C5L&C5R}). 
Counting the CKM suppression factors and the Yukawa coupling factors, 
it is easily shown that 
the $RRRR$ contribution to
the four-fermion operators $(u_R d_R)(s_L \nu_{\tau L})$
and $(u_R s_R)(d_L \nu_{\tau L})$ is dominated by a single 
(Higgsino dressing) diagram
exchanging $ \tilde{t}_R$ (the right-handed scalar top quark) 
and $ \tilde{\tau}_R $ (the right-handed scalar tau lepton). 
For $K^+\overline{\nu}_\mu$ and $K^+\overline{\nu}_e$,
the $RRRR$ contribution is negligible, 
since it is impossible to get a large Yukawa coupling
of the third-generation without small CKM suppression
factors in this case. 
The $LLLL$ contribution to $(u_L d_L)(s_L \nu_{i L})$
and $(u_L s_L)(d_L \nu_{i L})$ consists of two 
classes of (wino dressing) diagrams; they are
$ \tilde{c}_L$ exchange diagrams and $ \tilde{t}_L$ exchange
diagrams \cite{NCA}. 
Neglecting all of various subleading effects, 
we can write the amplitudes
(the coefficients 
of the four-fermion operators) for 
$p\rightarrow K^+ \overline{\nu}_i$ as, 
\begin{eqnarray}
{\rm Amp.}(p\rightarrow K^+ \overline{\nu}_e) & \sim & 
[ P_2 A_e(\tilde{c}_L) +
  P_3 A_e(\tilde{t}_L) ]_{LLLL}, \nonumber \\
{\rm Amp.}(p\rightarrow K^+ \overline{\nu}_\mu) & \sim & 
[ P_2 A_\mu(\tilde{c}_L) +
  P_3 A_\mu(\tilde{t}_L) ]_{LLLL}, \nonumber \\
{\rm Amp.}(p\rightarrow K^+ \overline{\nu}_\tau) & \sim & 
[ P_2 A_\tau(\tilde{c}_L) +
  P_3 A_\tau(\tilde{t}_L) ]_{LLLL}
+ [ P_1 A_\tau(\tilde{t}_R) ]_{RRRR}, 
\label{eqn:rough_estimation}
\end{eqnarray}
where the subscript $LLLL$ ($RRRR$) represents the contribution
from the $LLLL$ ($RRRR$) operator. 
We estimate $A_i$ by only the $(u d)(s \nu)$ type contributions 
here for simplicity, ignoring the $(u s)(d \nu)$ type contributions. 
The $LLLL$ contributions for $A_\tau$ 
can be written in a rough approximation as 
$A_\tau(\tilde{c}_L)$ $\sim$ $g_2^2 Y_c Y_b 
V_{ub}^* V_{cd}V_{cs}M_2$
$\!\! /(M_C m_{\tilde{f}}^2)$ and 
$A_\tau(\tilde{t}_L)$ $\sim$ $g_2^2 Y_t Y_b
V_{ub}^* V_{td}V_{ts}M_2$ 
$\!\! /(M_C m_{\tilde{f}}^2)$,
where $g_2$ is the weak SU(2) gauge coupling, and 
$M_2$ is a mass of the wino. 
A typical mass scale of the squarks and the sleptons 
is denoted by $m_{\tilde{f}}$. 
For $A_\mu$ and $A_e$, we just replace $Y_b V_{ub}^*$ in the
expressions for $A_\tau$ by $Y_s V_{us}^*$ and $Y_d V_{ud}^*$,
respectively. 
The $RRRR$ contribution is also evaluated as 
$A_\tau(\tilde{t}_R)$ $\sim$
$Y_d Y_t^2 Y_\tau V_{tb}^* V_{ud}V_{ts}\mu$
$\!\! /(M_C m_{\tilde{f}}^2)$,
where $\mu$ is the supersymmetric Higgsino mass. 
The magnitude of $\mu$ is determined from the 
radiative electroweak symmetry breaking condition, 
and satisfies $|\mu|$ $>$ $|M_2|$ 
in the present scenario. 

Relative magnitudes between these contributions are 
evaluated as follows. 
The magnitude of the $\tilde{c}_L$ contribution 
is comparable with that of the $\tilde{t}_L$ contribution 
for each generation mode: 
$|A_i(\tilde{c}_L)|$ $\sim$ $|A_i(\tilde{t}_L)|$. 
Therefore, cancellations between 
the $LLLL$ contributions 
$P_2 A_i(\tilde{c}_L)$ and $P_3 A_i(\tilde{t}_L)$ 
can occur simultaneously for three modes 
$p\rightarrow K^+\overline{\nu}_i$ ($i$ $=$ $e$, $\mu$ and $\tau$) 
by adjusting the relative phase $\phi_{23}$ 
between $P_2$ and $P_3$ \cite{NCA}. 
The magnitudes of the $LLLL$ contributions satisfy
$|P_2 A_\mu(\tilde{c}_L) + P_3 A_\mu(\tilde{t}_L)|$ 
$>$ $|P_2 A_\tau(\tilde{c}_L) + P_3 A_\tau(\tilde{t}_L)|$
$>$ $|P_2 A_e(\tilde{c}_L) + P_3 A_e(\tilde{t}_L)|$ 
independent of $\phi_{23}$. 
On the other hand, 
the magnitude of $A_\tau(\tilde{t}_R)$ is larger than 
those of $A_i(\tilde{c}_L)$ and $A_i(\tilde{t}_L)$, 
and the phase dependence of $P_1 A_\tau(\tilde{t}_R)$ is 
different from those of 
$P_2 A_i(\tilde{c}_L)$ and $P_3 A_i(\tilde{t}_L)$. 
Note that $A_i(\tilde{c}_L)$ and $A_i(\tilde{t}_L)$
are proportional to $\sim$ $1/(\sin \beta \cos \beta)$ $=$ 
$\tan \beta +  1/\tan \beta$, 
while $A_\tau(\tilde{t}_R)$ 
is proportional to $\sim$ $(\tan \beta +  1/\tan \beta)^2$,
where $\tan \beta$ is the ratio of the vacuum expectation values 
of the two Higgs bosons.
Hence the $RRRR$ contribution is more enhanced than the $LLLL$ 
contributions for large $\tan \beta$ \cite{RRRR}.

%
%
%
%
\section{Numerical results}
%
\hspace*{0.5cm}
%
%
Now we present the results of our numerical calculations. 
For the CKM matrix 
we adopt the standard parametrization \cite{PDG}, and
we fix the parameters as 
$V_{us}=0.2196$, $V_{cb}=0.0395$, $|V_{ub}/V_{cb}|=0.08$
and $\delta_{13}=90^\circ$ in the whole analysis,
where $\delta_{13}$ is a complex phase in the CKM matrix. 
The top quark mass is taken to be 175 $\gev$ \cite{m_top}. 
The colored Higgs mass $M_C$ and the heavy gauge boson mass
$M_V$ are assumed as $M_C$ $=$ $M_V$ $=$ $2 \times 10^{16} \gev$. 
We require 
constraint on $ b \rightarrow s \gamma $ branching ratio 
from CLEO \cite{CLEO} 
and bounds on SUSY particle masses obtained from 
direct searches at LEP \cite{Neutralino_Bound}, 
LEP II \cite{LEP2} and Tevatron \cite{CDF+D0}. 
We also impose condition to avoid color and charge breaking
vacua which is given in Ref.~\cite{CCB} at the electroweak scale. 

\begin{figure}[p]
\vspace*{-10mm}
\hspace*{-18mm}
\leavevmode\epsffile{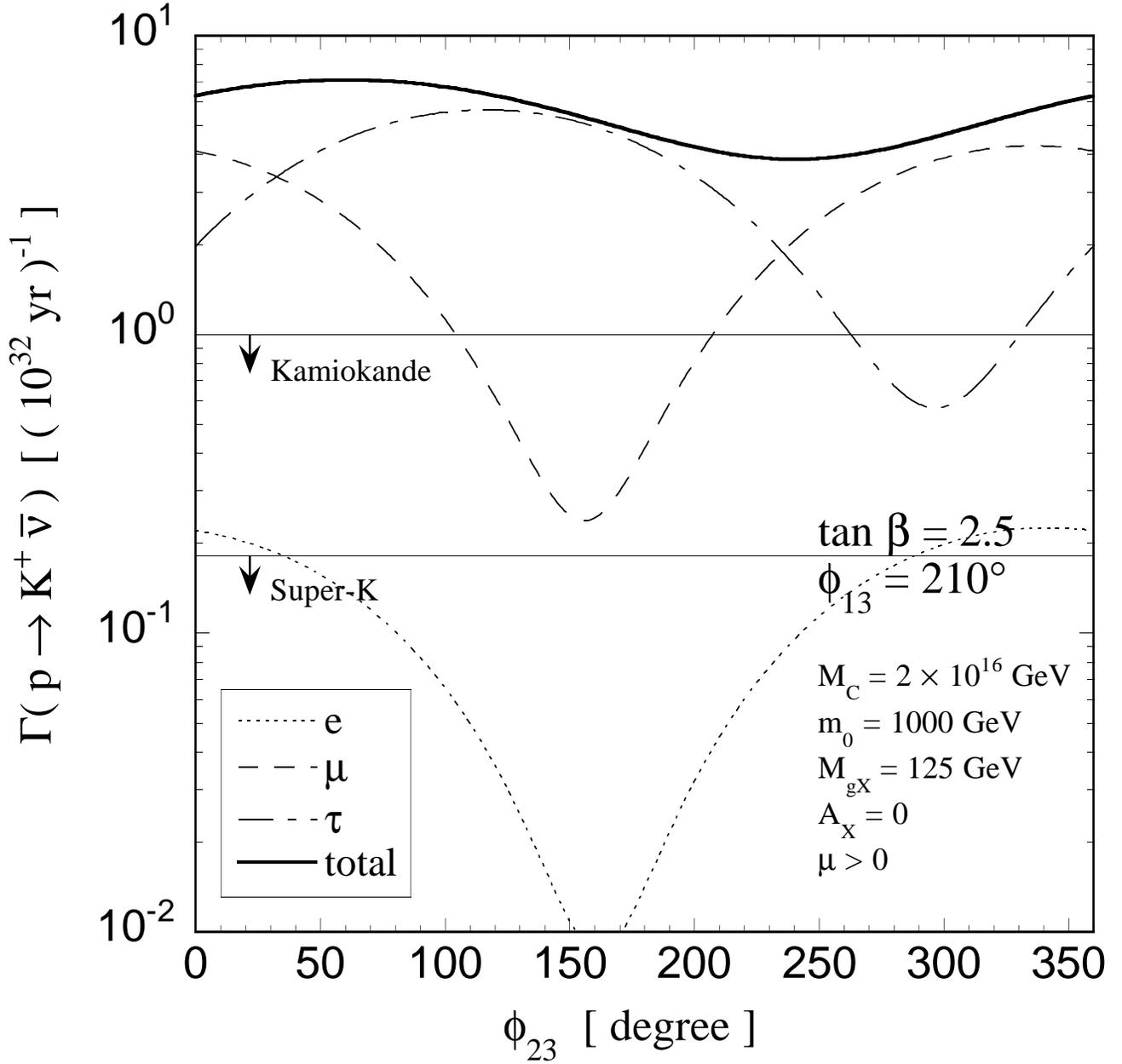}
\caption[figII]{Decay rates $\Gamma(p\rightarrow K^+ \overline{\nu}_i)$
($i$ $=$ $e$, $\mu$ and $\tau$) 
as functions of the phase $\phi_{23}$ for $\tan \beta$ $=2.5$. 
The other phase $\phi_{13}$ is fixed at $210^\circ$.
The CKM phase is taken as $\delta_{13}=90^\circ$. 
We fix the soft SUSY breaking parameters
as $m_0$ $=$ $1 \tev$, $M_{gX}$ $=$ $125 \gev$ and $A_X$ $=0$.
The sign of the supersymmetric Higgsino mass $\mu$ 
is taken to be positive. 
The colored Higgs mass $M_C$ and the heavy gauge boson mass
$M_V$ are assumed as $M_C$ $=$ $M_V$ $=$ $2 \times 10^{16} \gev$.
The horizontal lower line corresponds to 
the Super-Kamiokande limit  
$\tau(p\rightarrow K^+ \overline{\nu})$
$>$ $5.5 \times 10^{32}$ years, 
and the horizontal upper line corresponds to the Kamiokande limit 
$\tau(p\rightarrow K^+ \overline{\nu})$ 
$>$ $1.0 \times 10^{32}$ years.}
\label{fig:phi23}
\end{figure}

We mainly discuss the main decay mode 
$p\rightarrow K^+\overline{\nu}$ in this paper. 
We first discuss the effects of the phases $\phi_{13}$ and $\phi_{23}$
parametrizing the matrix $P$ in Eq.~(\ref{eqn:Phase_Matrix}).
In Fig.~\ref{fig:phi23} we present the dependence of 
the decay rates $\Gamma(p\rightarrow K^+ \overline{\nu}_i)$ 
on the phase $\phi_{23}$.  
As an illustration we fix the other phase $\phi_{13}$ at $210^\circ$, 
and later we consider 
the whole parameter space of $\phi_{13}$ and $\phi_{23}$. 
The soft SUSY breaking parameters are also fixed 
as $m_0$ $=$ $1 \tev$, $M_{gX}$ $=$ $125 \gev$ and $A_X$ $=0$ here.
The sign of the Higgsino mass $\mu$ 
is taken to be positive. 
With these parameters, 
all the masses of the scalar fermions 
other than the lighter $\tilde{t}$ are around 1 $\tev$, 
and the mass of the lighter $\tilde{t}$ is about 400 $\gev$. 
The lighter chargino is wino-like with a mass about 100 $\gev$. 
This figure shows that there is no region for 
the total decay rate $\Gamma(p\rightarrow K^+ \overline{\nu})$
to be strongly suppressed, thus the whole region of $\phi_{23}$ 
in Fig.~\ref{fig:phi23} is excluded by the Super-Kamiokande limit.
The phase dependence of 
$\Gamma(p\rightarrow K^+ \overline{\nu}_\tau)$ is 
quite different from those of 
$\Gamma(p\rightarrow K^+ \overline{\nu}_\mu)$ 
and $\Gamma(p\rightarrow K^+ \overline{\nu}_e)$. 
Though $\Gamma(p\rightarrow K^+ \overline{\nu}_\mu)$ 
and $\Gamma(p\rightarrow K^+ \overline{\nu}_e)$
are highly suppressed around $\phi_{23}$ $\sim$ $160^\circ$, 
$\Gamma(p\rightarrow K^+ \overline{\nu}_\tau)$ is not so 
in this region. 
There exists also the region $\phi_{23}$ $\sim$ $300^\circ$ where 
$\Gamma(p\rightarrow K^+ \overline{\nu}_\tau)$ is reduced. 
In this region, however, $\Gamma(p\rightarrow K^+ \overline{\nu}_\mu)$ 
and $\Gamma(p\rightarrow K^+ \overline{\nu}_e)$ 
are not suppressed in turn. 
Note also that the $K^+\overline{\nu}_\tau$ mode
can give the largest contribution.

\begin{figure}[p]
\vspace*{-10mm}
\hspace*{-18mm}
\leavevmode\epsffile{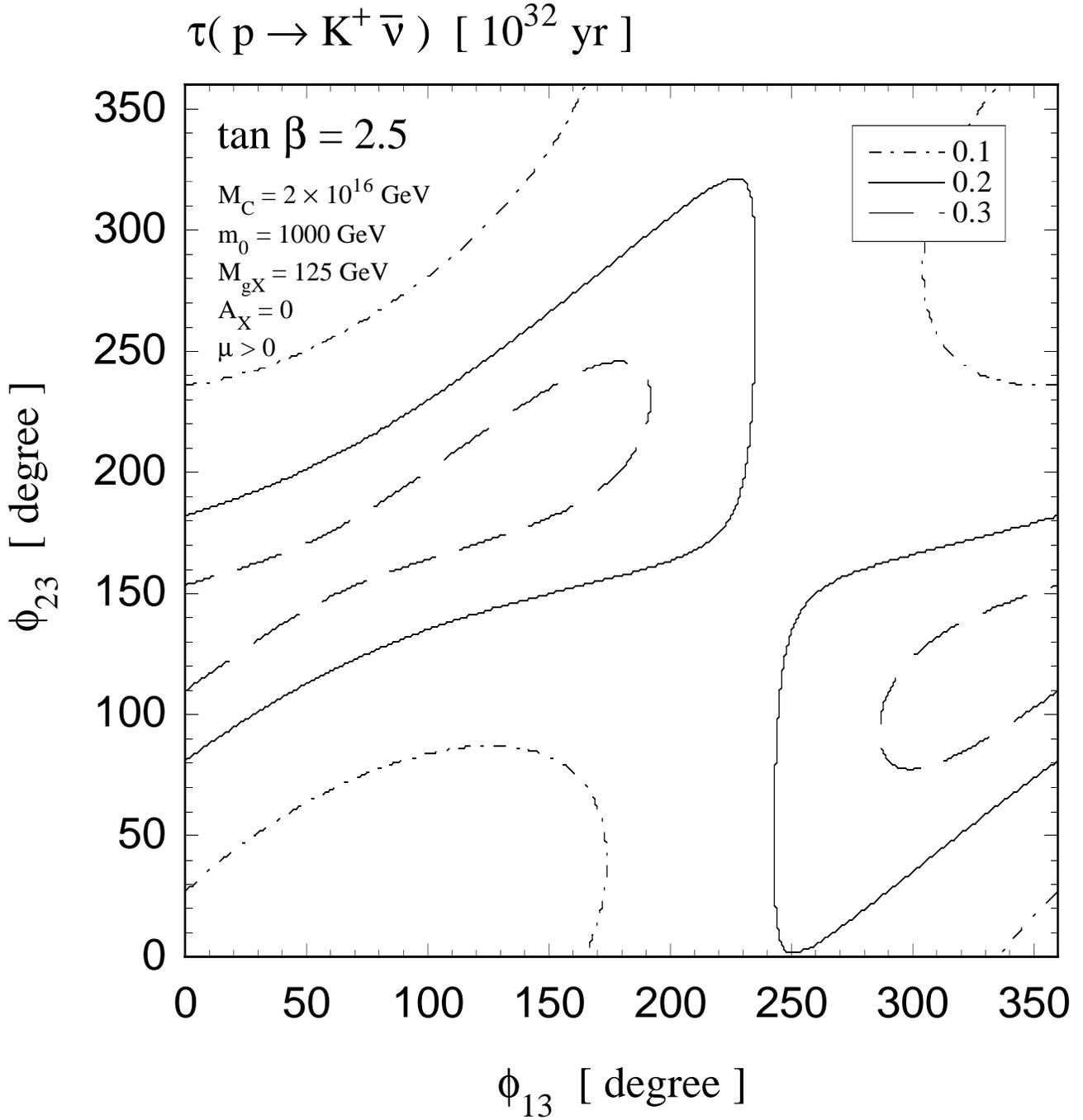}
\caption[figIII]{Contour plot for the partial lifetime
$\tau(p\rightarrow K^+ \overline{\nu})$ 
in the $\phi_{13}$-$\phi_{23}$ plane. The contributions of
three modes $K^+ \overline{\nu}_e$, $K^+ \overline{\nu}_\mu$ and
$K^+ \overline{\nu}_\tau$ are included. 
We use the same parameters as that in Fig.~\ref{fig:phi23}.
The maximum value of the contour is less than 
$0.5 \times 10^{32}$ years.}
\label{fig:contour_all}
\end{figure}

This behavior can be understood as follows. 
For $\overline{\nu}_\mu$ and $\overline{\nu}_e$, the effect of
the $RRRR$ operator is negligible, and 
the cancellation between the $LLLL$ contributions 
directly leads to the suppression of the decay rates. 
This cancellation indeed occurs around 
$\phi_{23}$ $\sim$ $160^\circ$ for 
both $\overline{\nu}_\mu$ 
and $\overline{\nu}_e$ simultaneously in Fig.~\ref{fig:phi23}. 
For $\overline{\nu}_\tau$, the situation is quite different.
The similar cancellation between 
$P_2 A_\tau(\tilde{c}_L)$ and $P_3 A_\tau(\tilde{t}_L)$ 
takes place around $\phi_{23}$ $\sim$ $160^\circ$ for 
$\overline{\nu}_\tau$ also. 
However, the $RRRR$ operator gives 
a significant contribution for $\overline{\nu}_\tau$. 
Therefore, $\Gamma(p\rightarrow K^+ \overline{\nu}_\tau)$
is not suppressed by the cancellation between the $LLLL$ contributions 
in the presence of the large $RRRR$ operator effect.
Notice that it is possible 
to reduce $\Gamma(p\rightarrow K^+\overline{\nu}_\tau)$ 
by another cancellation between the $LLLL$ contributions 
and the $RRRR$ contribution. 
This reduction of 
$\Gamma(p\rightarrow K^+\overline{\nu}_\tau)$ 
indeed appears around 
$\phi_{23}$ $\sim$ $300^\circ$ in Fig.~\ref{fig:phi23}. 
The decay rate $\Gamma(p\rightarrow K^+\overline{\nu}_\mu)$ is 
rather large in this region. 
The reason is that 
$P_2 A_\tau(\tilde{c}_L)$ and $P_3 A_\tau(\tilde{t}_L)$ 
are constructive in this 
region in order to cooperate with each other 
to cancel the large $RRRR$ contribution 
$P_1 A_\tau(\tilde{t}_R)$, hence 
$P_2 A_\mu(\tilde{c}_L)$ and $P_3 A_\mu(\tilde{t}_L)$
are also constructive in this region. 
Thus we cannot reduce both $\Gamma(p\rightarrow K^+\overline{\nu}_\tau)$
and $\Gamma(p\rightarrow K^+\overline{\nu}_\mu)$ simultaneously. 
Consequently, there is no region for 
the total decay rate $\Gamma(p\rightarrow K^+ \overline{\nu})$
to be strongly suppressed. 
%
In the previous analysis \cite{HMTY} the region 
$\phi_{23}$ $\sim$ $160^\circ$ has been considered to be allowed 
by the Kamiokande limit $\tau(p\rightarrow K^+ \overline{\nu})$ 
$>$ $1.0 \times 10^{32}$ years (90\,\% C.L.) \cite{Kam}. 
However the inclusion of the Higgsino dressing of the 
$RRRR$ operator excludes this region. 
%
In Fig.~\ref{fig:contour_all} we show a contour plot 
of the partial lifetime  
$\tau(p\rightarrow K^+\overline{\nu})$ 
in the $\phi_{13}$-$\phi_{23}$ plane.
It is found that there is no region to make
$\tau(p\rightarrow K^+ \overline{\nu})$ longer than 
$0.5 \times 10^{32}$ years. 
This implies that we cannot reduce 
both $\Gamma(p\rightarrow K^+\overline{\nu}_\tau)$
and $\Gamma(p\rightarrow K^+\overline{\nu}_\mu)$ simultaneously, 
even if we adjust the two phases $\phi_{13}$ and $\phi_{23}$ 
anywhere. 
Consequently, the whole parameter region in this plane 
is excluded by the Super-Kamiokande result.

\begin{figure}[p]
\vspace*{-10mm}
\hspace*{-18mm}
\leavevmode\epsffile{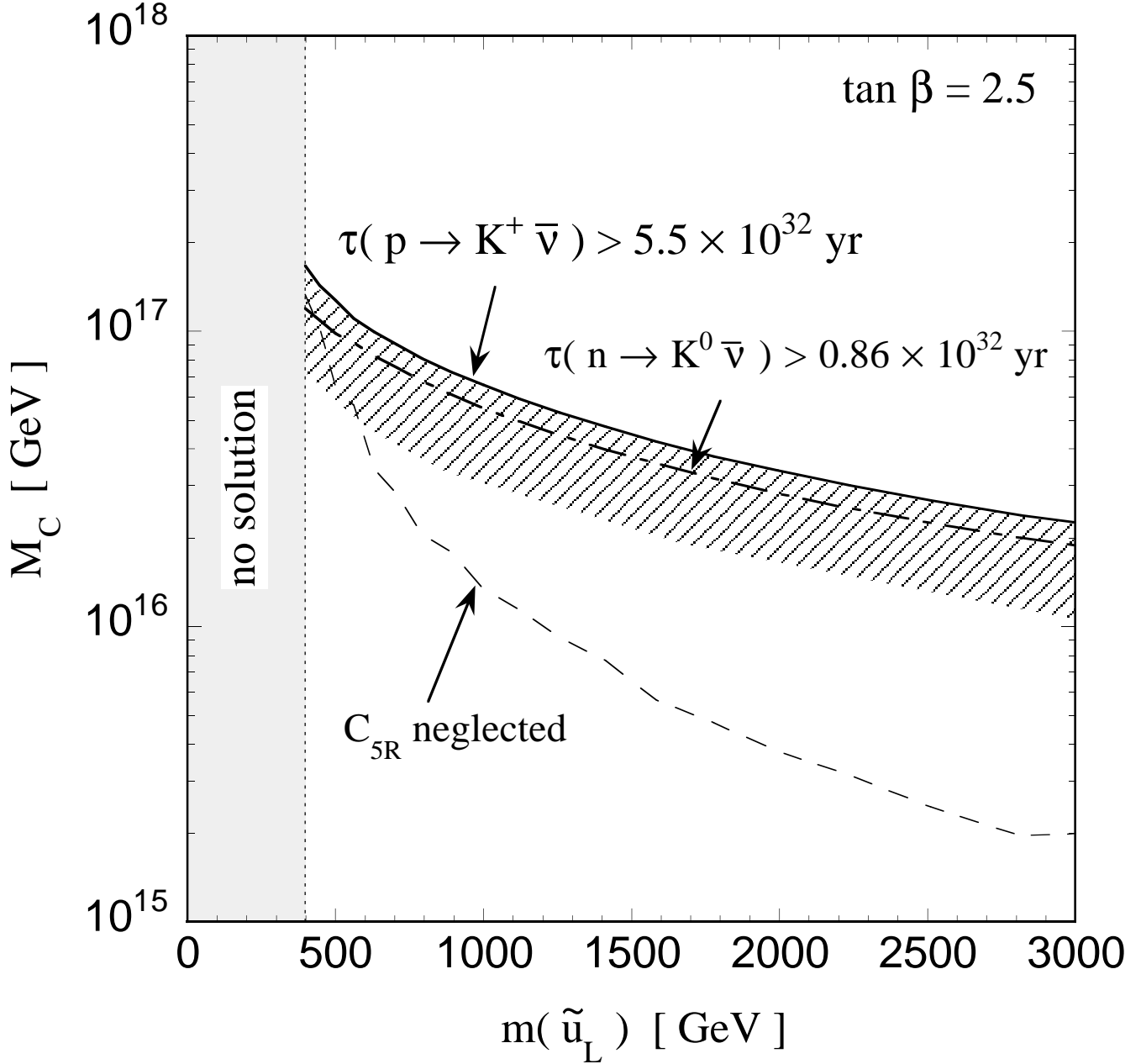}
\caption[figIV]{Lower bound on the colored Higgs mass $M_C$ 
as a function of the 
left-handed scalar up-quark mass $m_{\tilde{u}_L}$. 
The soft breaking parameters $m_0$, $M_{gX}$ and $A_X$ are scanned 
within the range of $0<m_0<3 \tev$, $0<M_{gX}<1 \tev$ and $ -5<A_X<5 $, 
and $\tan \beta$ is fixed at 2.5.
Both signs of $\mu$ are considered. 
The whole parameter region of the two phases $\phi_{13}$ and $\phi_{23}$ 
is examined. 
The solid curve represents the bound derived from the 
Super-Kamiokande limit $\tau(p\rightarrow K^+ \overline{\nu})$
$>$ $5.5 \times 10^{32}$ years, and 
the dashed curve represents the corresponding result 
without the $RRRR$ effect. 
Left-hand side of the vertical dotted line is excluded by 
other experimental constraints.
The dash-dotted curve represents the bound 
derived from the Kamiokande limit on the neutron partial lifetime 
$\tau (n\rightarrow K^0 \overline{\nu})$ 
$>$ $0.86 \times 10^{32}$ years.}
\label{fig:m_sf}
\end{figure}

Next we would like to consider the case where 
we vary the parameters we have fixed so far. 
The relevant parameters are 
the colored Higgs mass $M_C$, 
the soft SUSY breaking parameters and $\tan \beta$. 
As for the constants $\alpha_p$ and $\beta_p$
in the hadronic matrix elements, we have chosen 
the smallest value \cite{beta_p}. 
Hence other choices of these constants lead to 
enhancement of the proton decay rate which corresponds to 
severer constraints on this model. 
The partial lifetime $\tau(p\rightarrow K^+ \overline{\nu})$ 
is proportional to $M_C^2$ in a very good approximation,
since this mode is dominated by the dimension 5 operators. 
Using this fact and the calculated value of 
$\tau(p\rightarrow K^+ \overline{\nu})$ for the fixed 
$M_C$ $=$ $2 \times 10^{16} \gev$, we can obtain the lower bound
on $M_C$ from the experimental lower limit 
on $\tau(p\rightarrow K^+ \overline{\nu})$. 
In Fig.~\ref{fig:m_sf}, we present 
the lower bound on $M_C$ obtained from 
the Super-Kamiokande limit as a function of the 
left-handed scalar up-quark mass $m_{\tilde{u}_L}$. 
Masses of the scalar fermions other than the lighter $\tilde{t}$ 
are almost degenerate with $m_{\tilde{u}_L}$. 
The soft breaking parameters $m_0$, $M_{gX}$ and $A_X$ are scanned 
within the range of $0<m_0<3 \tev$, $0<M_{gX}<1 \tev$ and $ -5<A_X<5 $, 
and $\tan \beta$ is fixed at 2.5.
Both signs of $\mu$ are considered. 
The whole parameter region of the two phases $\phi_{13}$ and $\phi_{23}$ 
is examined. 
The solid curve in this figure 
represents the result with all the $LLLL$ and $RRRR$ contributions. 
It is shown that the lower bound on $M_C$ decreases 
like $\sim$ $1/m_{\tilde{u}_L}$ as $m_{\tilde{u}_L}$ increases. 
This indicates that the $RRRR$ effect is indeed relevant,
since the decay amplitude from the $RRRR$ operator is roughly proportional
to $\mu/(M_C m_{\tilde{f}}^2)$ $\sim$ $1/(M_C m_{\tilde{f}})$, 
where we use the fact that the magnitude of $\mu$ is determined from the
radiative electroweak symmetry breaking condition and
scales like $\mu$ $\sim$ $m_{\tilde{f}}$. 
The dashed curve in Fig.~\ref{fig:m_sf} 
represents the result in the case 
where we ignore the $RRRR$ effect. 
In this case 
the lower bound on $M_C$ behaves as 
$\sim$ $1/m_{\tilde{u}_L}^2$, since the $LLLL$ contribution 
is proportional to $M_2/(M_C m_{\tilde{f}}^2)$. 

It is found from the solid curve in Fig.~\ref{fig:m_sf}
that the colored Higgs mass $M_C$ must be larger 
than $6.5 \times 10^{16} \gev$ for $\tan \beta$ $=$ 2.5 when 
the typical sfermion mass is less than $1 \tev$. 
On the other hand, 
it has been pointed out that 
there exists an upper bound on $M_C$ given by 
$M_C$ $\leq$ $2.5 \times 10^{16} \gev$ (90\,\% C.L.) 
if we require the gauge coupling unification in the minimal 
contents of GUT superfields \cite{HMTY}. 
This upper bound is smaller than the lower bound derived from
our proton decay analysis.  
Therefore it turns out that 
the minimal SU(5) SUGRA GUT model 
with the sfermion masses less than $1 \tev$ 
is excluded for $\tan \beta$ $=2.5$. 
Note that the inclusion of the $RRRR$ effect is essential here. 
If we ignored the $RRRR$ effect, 
we could find allowed region around 
$1.2 \times 10^{16} \gev$ $\lsim$ 
$M_C$ $\lsim$ $2.5 \times 10^{16} \gev$. 
We can also see from Fig.~\ref{fig:m_sf} that 
the typical sfermion mass $m_{\tilde{f}}$ must be larger 
than about $2.5 \tev$ when $M_C$ is less than $2.5 \times 10^{16} \gev$ 
in the $\tan \beta$ $=2.5$ case. 
The $RRRR$ effect plays an essential role again, 
since the lower bound on $m_{\tilde{f}}$ would be $700 \gev$ 
if the $RRRR$ effect were ignored. 
We also find that 
the Kamiokande limit on the neutron partial lifetime 
$\tau (n\rightarrow K^0 \overline{\nu})$ 
$>$ $0.86 \times 10^{32}$ years (90\,\% C.L.) \cite{Kam} 
already gives a comparable bound with that derived here from 
the Super-Kamiokande limit on 
$\tau (p\rightarrow K^+ \overline{\nu})$, 
as shown by the dash-dotted curve in Fig.~\ref{fig:m_sf}. 
If the Super-Kamiokande updates the neutron limit  
from the Kamiokande, 
for example, by factor 5,  
then the lower bound on $M_C$ will become $\sqrt{5}$ times larger 
than that derived from the Kamiokande limit.

\begin{figure}[p]
\vspace*{-10mm}
\hspace*{-18mm}
\leavevmode\epsffile{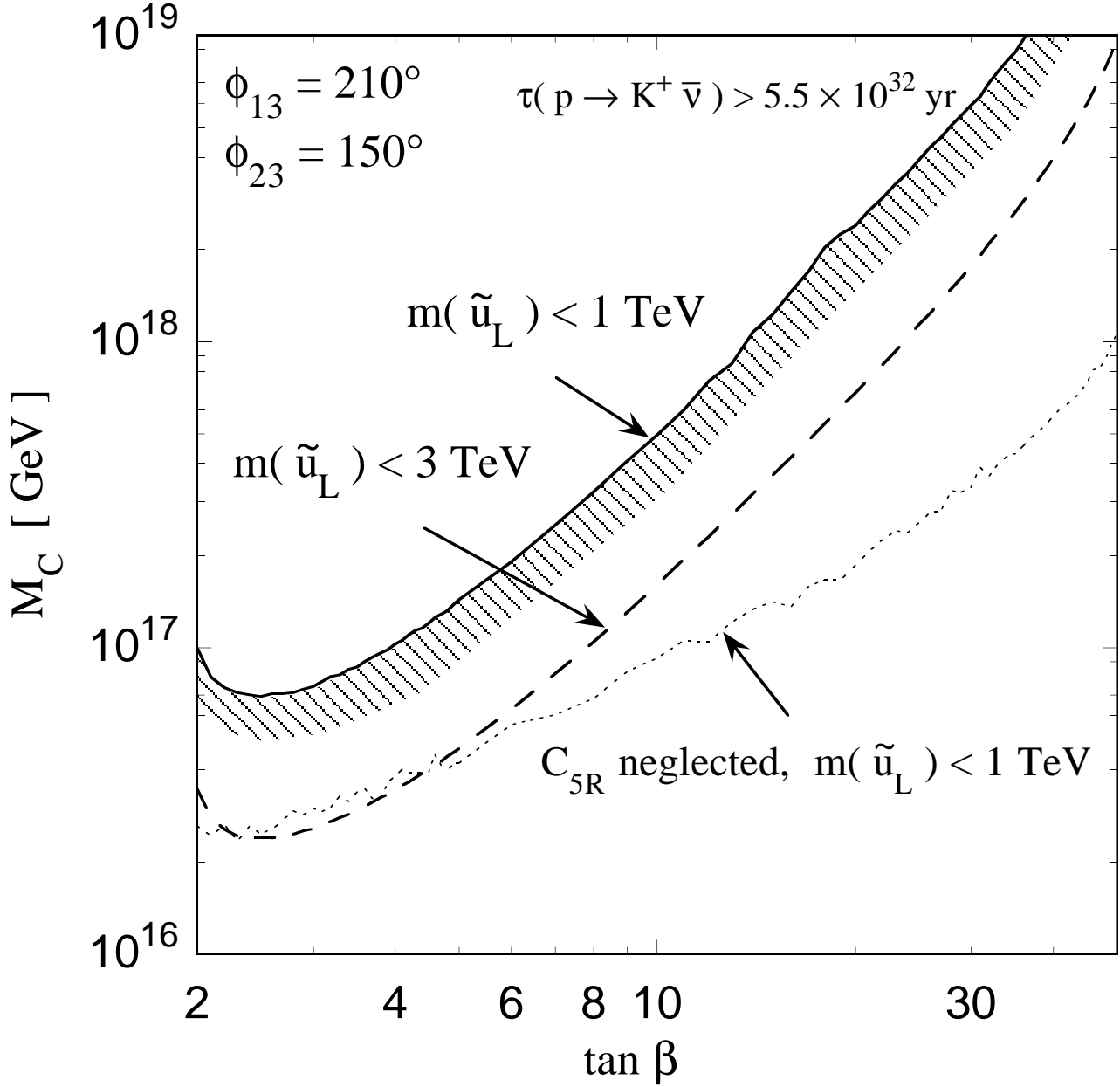}
\caption[figV]{The lower bound 
on the colored Higgs mass $M_C$ obtained from the Super-Kamiokande limit 
as a function of $\tan \beta$. 
The phase matrix $P$ is fixed by $\phi_{13}$ $=210^\circ$ 
and $\phi_{23}$ $=150^\circ$.
The region below the solid curve is excluded if the 
left-handed scalar up-quark mass $m_{\tilde{u}_L}$ is less than $1 \tev$. 
The lower bound reduces to the dashed curve if we 
allow $m_{\tilde{u}_L}$ up to $3 \tev$.
The result in the case where we ignore the $RRRR$ effect 
is shown by the dotted curve for $m_{\tilde{u}_L}$ $<$ $1 \tev$.}
\label{fig:tanB}
\end{figure}

Let us discuss the $\tan \beta$ dependence. 
Fig.~\ref{fig:tanB} shows the lower bound on the colored Higgs 
mass $M_C$ obtained from the Super-Kamiokande limit    
as a function of $\tan \beta$. 
Here we vary $m_0$, $M_{gX}$, $A_X$ and sign($\mu$) 
as in Fig.~\ref{fig:m_sf}.
The phases $\phi_{13}$ and $\phi_{23}$ are fixed as 
$\phi_{13}$ $=210^\circ$ and $\phi_{23}$ $=150^\circ$. 
The result does not change much even if we take other values of 
$\phi_{13}$ and $\phi_{23}$. 
The region below the solid curve is excluded if 
$m_{\tilde{u}_L}$ is less than $1 \tev$. 
The lower bound reduces to the dashed curve if we 
allow $m_{\tilde{u}_L}$ up to $3 \tev$. 
It is shown that the lower bound on $M_C$ 
behaves as $\sim$ $\tan^2 \beta$ in 
a large $\tan \beta$ region, as expected from the fact that 
the amplitude of $p\rightarrow K^+ \overline{\nu}_\tau$ 
from the $RRRR$ operator is roughly 
proportional to $\sim$ $\tan^2 \beta /M_C$. 
On the other hand the $LLLL$ contribution is 
proportional to $\sim$ $\tan \beta /M_C$, 
as shown by the dotted curve in Fig.~\ref{fig:tanB}. 
Thus the $RRRR$ operator is dominant for large $\tan \beta$ \cite{RRRR}. 
Note that the lower bound on $M_C$ has the minimum at 
$\tan \beta$ $\approx$ 2.5. 
Thus we can conclude that for other value of $\tan \beta$ 
the constraints on $M_C$ and $m_{\tilde{f}}$ become severer 
than those shown in Fig.~\ref{fig:m_sf}.

Finally, we comment on the other decay modes. 
For $p\rightarrow \pi^+\overline{\nu}$, we obtain a similar 
result with that for the $p\rightarrow K^+\overline{\nu}$ mode: 
the third-generation mode
$p\rightarrow \pi^+ \overline{\nu}_\tau$ is dominated by
the $RRRR$ effect, while the $RRRR$ effect is negligible 
for the first and the second generation modes. 
Let us define $r_i$ $=$ 
$\Gamma(p\rightarrow \pi^+ \overline{\nu}_i)$ 
$\!\!/ \Gamma(p\rightarrow K^+ \overline{\nu}_i)$ 
for $i$ $=$ $e$, $\mu$ and $\tau$. 
We see that $r_\mu$ $>1$ is realized in a part of the 
$\phi_{13}$-$\phi_{23}$ parameter region where 
$p\rightarrow K^+ \overline{\nu}_\mu$ mode is suppressed 
due to the cancellation between the $LLLL$ contributions. 
This result is consistent with that given in the previous 
analysis \cite{NCA}. 
As for the $\overline{\nu}_\tau$ mode, $r_\tau$ $>1$ is also possible 
in a different region where 
$\Gamma(p\rightarrow K^+ \overline{\nu}_\tau)$ is reduced. 
Consequently the ratio $r$ $=$ 
$\{ \sum_i \Gamma(p\rightarrow \pi^+ \overline{\nu}_i) \}$ 
$\!\!/ \{ \sum_i \Gamma(p\rightarrow K^+ \overline{\nu}_i) \}$ is 
smaller than 1 in the whole region of the 
$\phi_{13}$-$\phi_{23}$ space. 
Moreover it has been reported that 
the lattice calculation of the hadronic matrix elements \cite{Lattice} 
gives a smaller value of the ratio 
$\langle\pi |{\cal O}_{\not{B}}|p\rangle $ 
$\!\! / \langle K|{\cal O}_{\not{B}}|p\rangle$ 
than the chiral Lagrangian estimation, 
where ${\cal O}_{\not{B}}$ denotes the baryon number violating operators. 
Hence it follows that the ratio $r$ is expected to be smaller 
when we use the lattice result for the hadronic matrix element.
For the charged lepton mode 
$p\rightarrow M \ell^{+}$ ($M =K^0, \pi^0, \eta$ and $\ell=e, \mu$), 
effect of the $RRRR$ operator is quite small, since we cannot 
have the tau lepton in the final state. 

%
%
%
\section{Conclusions}
%
\hspace*{0.5cm}
We have reanalyzed the proton decay 
including the $RRRR$ dimension 5 operator 
in the minimal SU(5) SUGRA GUT model.
We have shown that the Higgsino dressing diagram of 
the $RRRR$ operator gives a dominant
contribution for 
$p\rightarrow K^+\overline{\nu}_\tau$,
and the decay rate of this mode 
can be comparable with that of 
$p\rightarrow K^+\overline{\nu}_\mu$. 
We have found that 
we cannot reduce both the decay rate of 
$p\rightarrow K^+\overline{\nu}_\tau$ and 
that of $p\rightarrow K^+\overline{\nu}_\mu$ simultaneously
by adjusting the relative phases $\phi_{13}$ and $\phi_{23}$ 
between the Yukawa couplings
at the colored Higgs interactions. 
We have obtained the bounds on the colored Higgs mass $M_C$ 
and the typical sfermion mass $m_{\tilde{f}}$ from 
the new limit on $\tau(p\rightarrow K^+ \overline{\nu})$ given by 
the Super-Kamiokande. 
The colored Higgs mass $M_C$ must be larger
than $6.5 \times 10^{16} \gev$ when 
$m_{\tilde{f}}$ is less than $1 \tev$. 
The typical sfermion mass $m_{\tilde{f}}$ must be larger than $2.5 \tev$ 
when $M_C$ is less than $2.5 \times 10^{16} \gev$.

%
%
%
\section*{Acknowledgements}
\hspace*{0.5cm} 
We would like to thank J. Hisano for a useful discussion, 
and Y. Okada for a careful reading of the manuscript. 
The work of T.G. was supported in part by the Soryushi Shogakukai.
The work of T.N. was supported in part by
the Grant-in-Aid for Scientific
Research from the Ministry of Education, Science and Culture, Japan. 
\appendix

\section{Formulas for the calculation of the nucleon decay}

In this appendix, we summarize the formulas used in the calculation of
the partial decay widths of the nucleon in the minimal SU(5) SUGRA GUT
in order to clarify our notations and conventions.
In the subsection \ref{sec:formulaMSSM}, generic formulas for the MSSM
are summarized.
The formulas specific to the calculation of the nucleon decay are given
in the subsection \ref{sec:formulapdecay}.

\subsection{MSSM part}
\label{sec:formulaMSSM}

\subsubsection{Superpotential}
\label{sec:MSSMsp}

Yukawa couplings of the Higgs doublets and matter fields and the
supersymmetric Higgs mass terms are given in the superpotential for the
MSSM which is written as
\begin{eqnarray}
  W_{\rm MSSM}
  &=& f_D^{ij} Q^\alpha_i D^c_j H_{1\alpha}
    + f_U^{ij} \epsilon_{\alpha\beta} Q^\alpha_i U^c_j H_2^\beta
    + f_L^{ij} \epsilon^{\alpha\beta} E^c_i L_{j\alpha} H_{1\beta}
    + \mu H_{1\alpha} H_2^\alpha
  \nonumber\\
  &=& f_D^{ij} \left( Q^u_i D^c_j H_1^- + Q^d_i D^c_j H_1^0 \right)
    + f_U^{ij} \left( Q^u_i U^c_j H_2^0 - Q^d_i U^c_j H_2^+ \right)
  \nonumber\\
  &&+ f_L^{ij} \left( E^c_i L^e_j H_1^0 - E^c_i L^\nu_j H_1^- \right)
    + \mu \left( H_1^0 H_2^0 + H_1^- H_2^+ \right) ~,
 \label{MSSMsuperpotential}
\end{eqnarray}
where $i,j$ and $\alpha,\beta$ are generation and SU(2) suffices,
respectively.
Color indices are suppressed for simplicity.
Components of the SU(2) doublets are denoted as
\begin{eqnarray}
  Q^\alpha_i 
  &=& \left( \begin{array}{c}  Q^u_i \\ Q^d_i \end{array}\right) ~,~~~
  L_{i\alpha}
  ~=~ \left( \begin{array}{cc} L^e_i & L^\nu_i \end{array}\right) ~,
  \nonumber\\
  H_{1\alpha}
  &=& \left( \begin{array}{cc} H_1^- & H_1^0 \end{array}\right) ~,~~~
  H_2^\alpha
  ~=~ \left( \begin{array}{c}  H_2^+ \\ H_2^0 \end{array}\right) ~.
  \label{doublets}
\end{eqnarray}
We take the generation basis for the superfields so that the Yukawa
coupling matrices (equivalently the mass matrices) for the up-type
quarks ($f_U$) and the leptons ($f_L$) should be diagonal (with real
positive diagonal elements) at the electroweak scale.
In this basis, the Yukawa coupling matrix for the down-type quarks $f_D$ 
is written as
\begin{eqnarray}
  f_D(m_Z) &=& V_{\rm KM}^* \hat{f}_D ~,
\end{eqnarray}
where $\hat{f}_D$ is diagonal (real positive) and $V_{\rm KM}$ is the
CKM matrix.
We take the PDG's ``standard'' phase convention for $V_{\rm KM}$
\cite{PDG}.
The SUSY Higgs mass parameter $\mu$ is taken as real in order to
automatically avoid a too-large electric dipole moments (EDMs) of the
neutron and the electron.
The sign of $\mu$ is taken as a free ``parameter''.

\subsubsection{Soft SUSY breaking terms}

Soft SUSY breaking terms of the MSSM are given as
\begin{eqnarray}
  -{\cal L}_{\rm soft}
  &=& (m_Q^2)^i_{~j} \wt{q}_i^\alpha \wt{q}^{\dagger j}_\alpha
    + (m_U^2)_i^{~j} \wt{u}^{\dagger i} \wt{u}_j
    + (m_D^2)_i^{~j} \wt{d}^{\dagger i} \wt{d}_j
  \nonumber\\
  &&+ (m_L^2)_i^{~j} \wt{l}^{\dagger i\alpha} \wt{l}_{j\alpha}
    + (m_E^2)^i_{~j} \wt{e}_i \wt{e}^{\dagger j} 
  \nonumber\\
  &&+ \Delta_1^2 h_1^{\dagger\alpha} h_{1\alpha}
    + \Delta_2^2 h_{2\alpha}^\dagger h_2^\alpha
    - \left( B\mu h_{1\alpha} h_2^\alpha + \hc \right)
  \nonumber\\
  &&+ \left(   A_U^{ij} \epsilon_{\alpha\beta}
                        \wt{q}^\alpha_i \wt{u}_j h_2^\beta
             + A_D^{ij} \wt{q}_i^\alpha \wt{d}_j h_{1\alpha}
             + A_L^{ij} \epsilon^{\alpha\beta}
                        \wt{e}_i \wt{l}_{j\alpha} h_{1\beta}
             + \hc \right)
  \nonumber\\
  &&+ \left(   \frac{M_1}{2} \wt{B}\wt{B}
             + \frac{M_2}{2} \wt{W}\wt{W}
             + \frac{M_3}{2} \wt{G}\wt{G} + \hc \right) ~.
  \label{MSSMsoft}
\end{eqnarray}
where $\wt{q}$, $\wt{d}$, $\wt{u}$, $\wt{e}$, $\wt{l}$, $h_1$ and $h_2$
are scalar components of $Q$, $D^c$, $U^c$, $E^c$, $L$, $H_1$ and $H_2$,
respectively, and $\wt{G}$, $\wt{W}$ and $\wt{B}$ are SU(3), SU(2) and
U(1) gaugino fields, respectively.
The gaugino masses $M_1$, $M_2$ and $M_3$ are taken as real positive.

In the minimal SUGRA GUT model, the soft SUSY breaking parameters
at the GUT scale $M_X$ are written in terms of the universal soft
SUSY breaking parameters $m_0$ (universal scalar mass), $M_{gX}$
(unified gaugino mass), and $A_X$ (dimensionless universal trilinear
coupling parameter):
\begin{subeqnarray}
  m_Q^2(M_X) &=& m_U^2(M_X) ~=~ m_D^2(M_X) ~=~ m_0^2{\bf 1} ~,
\\
  m_L^2(M_X) &=& m_E^2(M_X) ~=~ m_0^2{\bf 1} ~,
\\
  \Delta_1^2(M_X) &=& \Delta_2^2(M_X) ~=~ m_0^2 ~,
\\
  M_1(M_X) &=& M_2(M_X) ~=~ M_3(M_X) ~=~ M_{gX} ~,
\\
  A_U(M_X) &=& A_X m_0 f_U ~, ~~~
  A_D(M_X) ~=~ A_X m_0 f_D ~,
\\
  A_L(M_X) &=& A_X m_0 f_L ~,
\end{subeqnarray}
where ${\bf 1}$ is a $3\times3$ unit matrix in the generation space.
We take $A_X$ as real (with either sign) to avoid large EDMs.

\subsubsection{Mass matrices}

Mass matrices for squarks and sleptons are given as follows.
\begin{itemize}
\item up-type squark:
\begin{subeqnarray}
  {\cal M}_{\wt{u}}^2 &=&
  \left(
  \begin{array}{cc}
    m^2_{LL}(\wt{u}) & m^2_{LR}(\wt{u}) \\
    m^2_{RL}(\wt{u}) & m^2_{RR}(\wt{u}) 
  \end{array}
  \right) ~,
\\
  m^2_{LL}(\wt{u}) &=&
  v^2 s_\beta^2 f_U f_U^\dagger
    + m_Q^2
    + m_Z^2 c_{2\beta}
      \left( \frac{1}{2} - \frac{2}{3}s_W^2 \right){\bf 1}
~,
\\
  m^2_{RR}(\wt{u}) &=&
  v^2 s_\beta^2 f_U^\dagger f_U
    + m_U^2
    + m_Z^2 c_{2\beta}
      \left( \frac{2}{3}s_W^2 \right){\bf 1}
~,
\\
  m^2_{LR}(\wt{u}) &=&
   \mu^* f_U v c_\beta + A_U v s_\beta
 ~,
\\
    m^2_{RL}(\wt{u}) &=& m^{2\dagger}_{LR}(\wt{u}) ~,
\end{subeqnarray}
\item down-type squark:
\begin{subeqnarray}  
  {\cal M}_{\wt{d}}^2 &=&
  \left(
  \begin{array}{cc}
    m^2_{LL}(\wt{d}) & m^2_{LR}(\wt{d}) \\
    m^2_{RL}(\wt{d}) & m^2_{RR}(\wt{d})
  \end{array}
  \right) ~,
\\
    m^2_{LL}(\wt{d}) &=&
      v^2 c_\beta^2 f_D f_D^\dagger
    + m_Q^2
    + m_Z^2 c_{2\beta}
      \left( - \frac{1}{2} + \frac{1}{3}s_W^2 \right){\bf 1}
   ~,
\\
    m^2_{RR}(\wt{d}) &=&
      v^2 c_\beta^2 f_D^\dagger f_D
    + m_D^2
    + m_Z^2 c_{2\beta}
      \left( - \frac{1}{3} s_W^2 \right){\bf 1}
 ~,
\\
    m^2_{LR}(\wt{d}) &=&
      \mu^* f_D v s_\beta + A_D v c_\beta ~,
\\
    m^2_{RL}(\wt{d}) &=& m^{2\dagger}_{LR}(\wt{d}) ~,
\end{subeqnarray}
\item charged slepton:
\begin{subeqnarray}
  {\cal M}^2_{\wt{l}} &=&
  \left(
  \begin{array}{cc}
    m^2_{LL}(\wt{l}) & m^2_{LR}(\wt{l}) \\
    m^2_{RL}(\wt{l}) & m^2_{RR}(\wt{l})
  \end{array}
  \right) ~,
\\
  m^2_{LL}(\wt{l}) &=&
  v^2 c_\beta^2 f_L^\dagger f_L
    + m_L^2
    + m_Z^2 c_{2\beta}
      \left( - \frac{1}{2} + s_W^2 \right)
      {\bf 1} ~,
\\
  m^2_{RR}(\wt{l}) &=&
  v^2 c_\beta^2 f_L f_L^\dagger
    + m_E^2
    + m_Z^2 c_{2\beta}
      \left( - s_W^2 \right)
      {\bf 1} ~,
\\
  m^2_{RL}(\wt{l}) &=&
  \mu^* f_L v s_\beta + A_L v c_\beta ~,
\\
    m^2_{LR}(\wt{l}) &=& m^{2\dagger}_{LR}(\wt{l}) ~,
\end{subeqnarray}
\item sneutrino:
\begin{subeqnarray}
  {\cal M}_{\wt{\nu}}^2 &=& 
    m_L^2
    + m_Z^2 c_\beta^2
      \left( \frac{1}{2} \right)
     {\bf 1} ~,
\end{subeqnarray}
\end{itemize}
where $c_\beta=\cos\beta>0$, $s_\beta=\sin\beta>0$,
$c_{2\beta}=\cos 2\beta$, $s_W=\sin\theta_W$ and
$v^2=\vev{h_1}^2+\vev{h_2}^2$ ($v \approx 174$ GeV).
The above mass matrices are diagonalized with use of 
$6\times6$ unitary matrices $\wt{U}_U$, $\wt{U}_D$ and $\wt{U}_L$,
and a $3\times3$ unitary matrix $\wt{U}_N$, which are defined as
\begin{subeqnarray}
  \wt{U}_U {\cal M}_{\wt{u}}^{2{\bf T}} \wt{U}_U^\dagger &=&
  \mbox{diagonal}(m^2_{\wt{u}_I}) ~,
\\
  \wt{U}_D {\cal M}_{\wt{d}}^{2{\bf T}} \wt{U}_D^\dagger &=&
  \mbox{diagonal}(m^2_{\wt{d}_I}) ~,
\\
  \wt{U}_L^\dagger {\cal M}^2_{\wt{l}} \wt{U}_L &=&
  \mbox{diagonal}(m^2_{\wt{l}_I}) ~,
\\
  \wt{U}_N^\dagger {\cal M}^2_{\wt{\nu}} \wt{U}_N &=&
  \mbox{diagonal}(m^2_{\wt{\nu}_i}) ~,
\end{subeqnarray}
where $^{\bf T}$ stands for the transpose.

Mass matrices for charginos (${\cal M}_C$) and neutralinos
(${\cal M}_N$) are given as follows.
\begin{subeqnarray}
    {\cal M}_C &=&
    \left( 
    \begin{array}{cc}
               M_2         & \sqrt{2}m_W s_\beta \\
      -\sqrt{2}m_W c_\beta &       -\mu
    \end{array}
  \right) ~,
\\
{\cal M}_N &=&
  \left(
    \begin{array}{cccc}
      -M_1 &  0   & -m_Z s_W c_\beta & \phantom{-}m_Z s_W s_\beta \\
       0   & -M_2 & \phantom{-}m_Z c_W c_\beta & -m_Z c_W s_\beta \\
      -m_Z s_W c_\beta &  \phantom{-}m_Z c_W c_\beta & 0 & \mu \\
      \phantom{-}m_Z s_W s_\beta & -m_Z c_W s_\beta & \mu & 0
    \end{array}
  \right) ~.
\end{subeqnarray}
${\cal M}_C$ and ${\cal M}_N$ are diagonalized with $2\times2$ unitary
matrices $U_\pm$ and a $4\times4$ unitary matrix $U_N$, respectively,
which are defined as
\begin{subeqnarray}
 -U_-^\dagger {\cal M}_C U_+ &=&
\mbox{diagonal}(M_C^\alpha) ~,
\\
  U_N^{\bf T} {\cal M}_N U_N &=& 
\mbox{diagonal}(M_N^{\ol{\alpha}}) ~,
\end{subeqnarray}
where all mass eigenvalues $M_C^\alpha$ ($\alpha=1,2$) and
$M_N^{\ol{\alpha}}$ ($\ol{\alpha}=1,2,3,4$) are taken as real positive.

\subsubsection{Interaction Lagrangian in mass basis}

The quark (lepton) -- squark (slepton) -- ino (gluino, chargino,
neutralino) interaction terms are given as follows.
\begin{subeqnarray}
{\cal L}_{\rm int} &=& 
  {\cal L}_{\rm int}(\wt{G})
+ {\cal L}_{\rm int}(\chi^\pm)
+ {\cal L}_{\rm int}(\chi^0) ~,
\nonumber\\
{\cal L}_{\rm int}(\wt{G}) &=&
  -i\sqrt{2} g_3 \wt{d}^{* I} \ol{\wt{G}} \left[
    \left( \Gamma_{GL}^{(d)} \right)_{I}^{j} \PL
  + \left( \Gamma_{GR}^{(d)} \right)_{I}^{j} \PR
  \right] d_j
\nonumber\\&&
  -i\sqrt{2} g_3 \wt{u}^{* I} \ol{\wt{G}} \left[
    \left( \Gamma_{GL}^{(u)} \right)_{I}^{j} \PL
  + \left( \Gamma_{GR}^{(u)} \right)_{I}^{j} \PR
  \right] u_j  + \hc ~,
\\
{\cal L}_{\rm int}(\chi^\pm) &=& \phantom{+}
  g_2 \ol{\chi}^-_\alpha \left[
    \left( \Gamma_{CL}^{(d)} \right)_{I}^{\alpha j} \PL
  + \left( \Gamma_{CR}^{(d)} \right)_{I}^{\alpha j} \PR
  \right] d_j \wt{u}^{* I}
\nonumber\\&& +
  g_2 \ol{\chi}^+_\alpha \left[
    \left( \Gamma_{CL}^{(u)} \right)_{I}^{\alpha j} \PL
  + \left( \Gamma_{CR}^{(u)} \right)_{I}^{\alpha j} \PR
  \right] u_j \wt{d}^{* I}
\nonumber\\&& +
  g_2 \ol{\chi}^-_\alpha \left[
    \left( \Gamma_{CL}^{(l)} \right)_{i}^{\alpha j} \PL
  + \left( \Gamma_{CR}^{(l)} \right)_{i}^{\alpha j} \PR
  \right] l_j \wt{\nu}^{* i}
\nonumber\\&& +
  g_2 \ol{\chi}^+_\alpha
    \left( \Gamma_{CL}^{(\nu)} \right)_{I}^{\alpha j} \PL
  \nu_j \wt{l}^{* I} + \hc ~,
\\
{\cal L}_{\rm int}(\chi^0) &=& \phantom{+}
  g_2 \ol{\chi}^0_{\ol{\alpha}} \left[
    \left( \Gamma_{NL}^{(d)} \right)_{I}^{\ol{\alpha} j} \PL
  + \left( \Gamma_{NR}^{(d)} \right)_{I}^{\ol{\alpha} j} \PR
  \right] d_j \wt{d}^{* I}
\nonumber\\&& +
  g_2 \ol{\chi}^0_{\ol{\alpha}} \left[
    \left( \Gamma_{NL}^{(u)} \right)_{I}^{\ol{\alpha} j} \PL
  + \left( \Gamma_{NR}^{(u)} \right)_{I}^{\ol{\alpha} j} \PR
  \right] u_j \wt{u}^{* I}
\nonumber\\&& +
  g_2 \ol{\chi}^0_{\ol{\alpha}} \left[
    \left( \Gamma_{NL}^{(l)} \right)_{I}^{\ol{\alpha} j} \PL
  + \left( \Gamma_{NR}^{(l)} \right)_{I}^{\ol{\alpha} j} \PR
  \right] l_j \wt{l}^{* I}
\nonumber\\&& +
  g_2 \ol{\chi}^0_{\ol{\alpha}}
    \left( \Gamma_{NL}^{(\nu)} \right)_{i}^{\ol{\alpha} j} \PL
  \nu_j \wt{\nu}^{* i} + \hc ~,
\end{subeqnarray}
where
$\PL = \frac{1}{2}(1-\gamma_5)$ and $\PR = \frac{1}{2}(1+\gamma_5)$, 
$g_2$ and $g_3$ are SU(2) and SU(3) gauge coupling constants,
respectively.
Here and hereafter,
$\wt{G}$, $\chi^\pm_\alpha$, $\chi^0_{\ol{\alpha}}$, $\wt{u}_I$,
$\wt{d}_I$, $\wt{l}_I$, $\wt{\nu}_i$, $u_i$, $d_i$, $l_i$ and $\nu_i$
denote gluino, chargino, neutralino, up-type squark, down-type squark,
charged slepton, sneutrino, up-type quark, down-type quark, charged
lepton and neutrino fields in mass basis, respectively.
Ranges of the suffices are $I = 1, 2, \cdots, 6$ (squarks and charged
sleptons), $i, j, k = 1, 2, 3$ (quarks, leptons and sneutrinos),
$\alpha = 1, 2$ (charginos) and $\ol{\alpha} = 1, 2, 3, 4$ (neutralinos).
Mixing factors at each vertex are written in terms of the
mass-diagonalizing matrices $\wt{U}_U$, $\wt{U}_D$, $\wt{U}_L$,
$\wt{U}_N$, $U_\pm$ and $U_N$ as follows.
\begin{itemize}
\item gluino:
\begin{subeqnarray}
    \left( \Gamma_{GL}^{(d)} \right)_{I}^{j} &=&
      \sum_{k=1}^3 \left( \wt{U}_D   \right)_I^{~k}
                   \left( V_{\rm KM} \right)_k^{~j} ~,
\\
    \left( \Gamma_{GR}^{(d)} \right)_{I}^{j} &=&
      \left( \wt{U}_D   \right)_I^{~j+3} ~,
\\
    \left( \Gamma_{GL}^{(u)} \right)_{I}^{j} &=&
      \left( \wt{U}_U   \right)_I^{~j} ~,
\\
    \left( \Gamma_{GR}^{(u)} \right)_{I}^{j} &=&
      \left( \wt{U}_U   \right)_I^{~j+3} ~,
\end{subeqnarray}
\item chargino:
\begin{subeqnarray}
    \left( \Gamma_{CL}^{(d)} \right)_{I}^{\alpha j} &=&
      \sum_{k=1}^3 \Biggl\{
        \left( \wt{U}_U \right)_I^{~k}
        \left( U_+ \right)_1^{~\alpha}
\nonumber\\&&
      + \left( \wt{U}_U \right)_I^{~k+3}
        \frac{m_k^{(u)}}{\sqrt{2}m_Ws_\beta}
        \left( U_+ \right)_2^{~\alpha}
      \Biggr\} \left( V_{\rm KM} \right)_k^{~j} ~,
\\
    \left( \Gamma_{CR}^{(d)} \right)_{I}^{\alpha j} &=&
     -\sum_{k=1}^3 
        \left( \wt{U}_U \right)_I^{~k}
        \left( V_{\rm KM} \right)_k^{~j}
        \frac{m_j^{(d)}}{\sqrt{2}m_Wc_\beta}
        \left( U_- \right)_2^{~\alpha} ~,
\\
    \left( \Gamma_{CL}^{(u)} \right)_{I}^{\alpha j} &=&
        \left( \wt{U}_D \right)_I^{~j}
        \left( U_-^\dagger \right)_\alpha^{~1}
\nonumber\\&& 
      - \sum_{k=1}^3
        \left( \wt{U}_D \right)_I^{~k+3}
        \frac{m_k^{(d)}}{\sqrt{2}m_Wc_\beta}
        \left( V_{\rm KM}^\dagger \right)_k^{~j}
        \left( U_-^\dagger \right)_\alpha^{~2} ~,
\\
    \left( \Gamma_{CR}^{(u)} \right)_{I}^{\alpha j} &=&
        \left( \wt{U}_D \right)_I^{~j}
        \frac{m_j^{(u)}}{\sqrt{2}m_Ws_\beta}
        \left( U_+^\dagger \right)_\alpha^{~2} ~,
\\
    \left( \Gamma_{CL}^{(l)} \right)_{i}^{\alpha j} &=&
     -\left( \wt{U}_N^\dagger \right)_i^{~j}
      \left( U_+ \right)_1^{~\alpha} ~,
\\
    \left( \Gamma_{CR}^{(l)} \right)_{i}^{\alpha j} &=&
      \frac{m_j^{(l)}}{\sqrt{2}m_Wc_\beta}
      \left( \wt{U}_N^\dagger \right)_i^{~j}
      \left( U_- \right)_2^{~\alpha} ~,
\\
    \left( \Gamma_{CL}^{(\nu)} \right)_{I}^{\alpha j} &=&
     -\left( \wt{U}_L^\dagger \right)_I^{~j}
      \left( U_-^\dagger \right)_{\alpha}^{~1}
\nonumber\\&& 
     +\frac{m_j^{(l)}}{\sqrt{2}m_Wc_\beta}
      \left( \wt{U}_L^\dagger \right)_{I}^{~j+3}
      \left( U_-^\dagger \right)_{\alpha}^{~2} ~,
\end{subeqnarray}
\item neutralino:
\begin{subeqnarray}
    \left( \Gamma_{NL}^{(d)} \right)_{I}^{\ol{\alpha} j} &=&
      \sqrt{2}\left[+\frac{1}{2} \left( U_N \right)_2^{~\ol{\alpha}}
      - \frac{1}{6} t_W \left( U_N \right)_1^{~\ol{\alpha}}
      \right] \sum_{k=1}^{3} \left( \wt{U}_{D} \right)_I^{~k}
              \left( V_{\rm KM} \right)_k^{~j}
\nonumber\\&&
    - \frac{m_j^{(d)}}{\sqrt{2}m_Wc_\beta} \left( U_N \right)_3^{~\ol{\alpha}}
      \left( \wt{U}_{D} \right)_I^{~j+3} ~,
\\
    \left( \Gamma_{NR}^{(d)} \right)_{I}^{\ol{\alpha} j} &=&
      \sqrt{2}\left[ 
      - \frac{1}{3} t_W \left( U_N^\dagger \right)_{\ol{\alpha}}^{~1}
      \right] \left( \wt{U}_{D} \right)_I^{~j+3}
\nonumber\\&&
    - \frac{m_j^{(d)}}{\sqrt{2}m_Wc_\beta}
      \left( U_N^\dagger \right)_{\ol{\alpha}}^{~3}
      \sum_{k=1}^3 \left( \wt{U}_{D} \right)_I^{~k}
                   \left( V_{\rm KM} \right)_k^{~j} ~,
\\
    \left( \Gamma_{NL}^{(u)} \right)_{I}^{\ol{\alpha} j} &=&
      \sqrt{2}\left[ -\frac{1}{2} \left( U_N \right)_2^{~\ol{\alpha}}
      - \frac{1}{6} t_W \left( U_N \right)_1^{~\ol{\alpha}}
      \right] \left( \wt{U}_{U} \right)_I^{~j}
\nonumber\\&&
    - \frac{m_j^{(u)}}{\sqrt{2}m_Ws_\beta}
      \left( U_N \right)_4^{~\ol{\alpha}}
      \left( \wt{U}_{U} \right)_I^{~j+3} ~,
\\
    \left( \Gamma_{NR}^{(u)} \right)_{I}^{\ol{\alpha} j} &=&
      \sqrt{2}\left[ 
      +\frac{2}{3} t_W \left( U_N^\dagger \right)_{\ol{\alpha}}^{~2}
      \right] \left( \wt{U}_{U} \right)_I^{~j+3}
\nonumber\\&&
    - \frac{m_j^{(u)}}{\sqrt{2}m_Ws_\beta}
      \left( U_N^\dagger \right)_{\ol{\alpha}}^{~4}
      \left( \wt{U}_{U} \right)_I^{~j} ~,
\\
    \left( \Gamma_{NL}^{(l)} \right)_{I}^{\ol{\alpha} j} &=&
      \sqrt{2}\left[ \frac{1}{2} \left( U_N \right)_2^{~\ol{\alpha}}
      + \frac{1}{2} t_W \left( U_N \right)_1^{~\ol{\alpha}}
      \right] \left( \wt{U}_{L}^\dagger \right)_I^{~j}
\nonumber\\&&
    - \frac{m_j^{(l)}}{\sqrt{2}m_Wc_\beta}
      \left( U_N \right)_{3}^{~\ol{\alpha}}
      \left( \wt{U}_{L}^\dagger \right)_I^{~j+3} ~,
\\
    \left( \Gamma_{NR}^{(l)} \right)_{I}^{\ol{\alpha} j} &=&
      \sqrt{2}\left[
        - t_W \left( U_N^\dagger \right)_{\ol{\alpha}}^{~1}
      \right] \left( \wt{U}_{L}^\dagger \right)_I^{~j+3}
\nonumber\\&&
    - \frac{m_j^{(l)}}{\sqrt{2}m_Wc_\beta}
      \left( U_N^\dagger \right)_{\ol{\alpha}}^{~3}
      \left( \wt{U}_{L}^\dagger \right)_I^{~j} ~,
\\
    \left( \Gamma_{NL}^{(\nu)} \right)_{i}^{\ol{\alpha} j} &=&
      \sqrt{2}\left[
        - \frac{1}{2} \left( U_N \right)_2^{~\ol{\alpha}}
        + \frac{1}{2} t_W \left( U_N \right)_1^{~\ol{\alpha}}
      \right] \left( \wt{U}_{N}^\dagger \right)_i^{~j} ~,
\end{subeqnarray}
\end{itemize}
where $t_W=\tan\theta_W$ and $m^{(u)}_i$, $m^{(d)}_i$ and $m^{(l)}_i$
are masses (real positive) of up-type quarks, down-type quarks and
charged leptons, respectively.

\subsection{Formulas specific to the nucleon decay}
\label{sec:formulapdecay}

\subsubsection{Dimension five operators}

Dimension five operators relevant to the nucleon decay are described by
the following superpotential:
\begin{eqnarray}
  W_5 &=& -\frac{1}{M_C}
  \left\{
    C_{5L}^{ijkl}
    \frac{1}{2}\epsilon_{\hat{a}\hat{b}\hat{c}} \epsilon_{\alpha\beta}
    Q_k^{\hat{a}\alpha} Q_l^{\hat{b}\beta} Q_i^{\hat{c}\gamma} L_{j\gamma}
  + C_{5R}^{ijkl}
    \epsilon^{\hat{a}\hat{b}\hat{c}}
    E^c_k U^c_{l\hat{a}} U^c_{i\hat{b}} D^c_{j\hat{c}}
  \right\} ~,
\end{eqnarray}
where the suffices $\hat{a},\hat{b},\hat{c}$ are color indices.
The coefficients $C_{5L}$ and $C_{5R}$ are given at the GUT scale in
terms of the Yukawa coupling matrices:
\begin{subeqnarray}
  C_{5L}^{ijkl}(M_X) &=&
  f_D^{im}(M_X)\, (V_{DL})_m^{~j}\,
  f_U^{kn}(M_X)\, (V_{QU}^\dagger)_n^{~l}
~,
\\
  C_{5R}^{ijkl}(M_X) &=&
  f_D^{mj}(M_X)\, (V_{QU})_m^{~i}\,
  f_U^{nl}(M_X)\, (V_{QE})_n^{~k}
~,  
\label{eq:C5LRatMX}
\end{subeqnarray}
where $V_{QU}$, $V_{QE}$ and $V_{DL}$ are $3\times3$ unitary matrices
which parametrize the differences between generation bases of the MSSM
superfields embedded in SU(5) superfields $\Psi(10)$ and $\Phi(\ol{5})$; 
\ie, the MSSM multiplets are accomodated into $\Psi$ and $\Phi$ as
\begin{subeqnarray}
  \Psi_i &\Leftarrow&
  \left\{
    Q_i ~,\
    ( V_{QU} )_{i}^{~k} U^c_k ~,\
    ( V_{QE} )_{i}^{~k} E^c_k
  \right\} ~,
\\
  \Phi_i &\Leftarrow&
  \left\{
    D^c_i ~,\
    ( V_{DL} )_{i}^{~k} L_k
  \right\} ~.
\end{subeqnarray}
$V_{QU}$, $V_{QE}$ and $V_{DL}$ are determined by the unitary matrices
which diagonalize the Yukawa coupling matrices at $M_X$, and the phase
matrix $P$:
\begin{subeqnarray}
  V_{QU} &=& U_Q^{(u)\dagger}P^\dagger U_U ~,
\\
  V_{QE} &=& U_Q^{(d)\dagger} U_E ~,
\\
  V_{DL} &=& U_D^\dagger U_L ~,
\end{subeqnarray}
where the Yukawa coupling matrices are diagonalized with $U$'s as
\begin{subeqnarray}
  U_Q^{(u)*}\,f_{U}(M_X)\,U_U^\dagger &=& Y_U ~,
\\
  U_Q^{(d)*}\,f_{D}(M_X)\,U_D^\dagger &=& Y_D ~,
\\
  U_E^*\,f_{L}(M_X)\,U_L^\dagger &=& Y_L ~.
\end{subeqnarray}
$Y_U$, $Y_D$ and $Y_L$ are diagonal matrices with real positive diagonal
elements.
The CKM matrix at the GUT scale $V\equiv V_{\rm KM}(M_X)$ is also
written in terms of $U$'s as
\begin{eqnarray}
  V &=& U_Q^{(u)}U_Q^{(d)\dagger} ~.
\end{eqnarray}
In the present genaration basis described in Sec.~\ref{sec:MSSMsp},
$U_Q^{(u)},U_U,U_D\approx{\bf 1}$, $U_Q^{(d)}\approx V_{\rm KM}^\dagger$
and $U_E=U_L={\bf 1}$.
Consequently,
\begin{eqnarray}
  V_{QU} &\approx& P^\dagger ~,
~~~
  V_{QE} ~\approx~ V_{\rm KM} ~\approx~ V ~,
~~~
  V_{DL} ~\approx~ {\bf 1} ~,
\end{eqnarray}
The expressions for $C_{5L,R}$ in Eq.~(\ref{eqn:C5L&C5R}) are obtained 
from Eq.~(\ref{eq:C5LRatMX}) in this approximation.

In the component form, the dimension five operators at the electroweak
scale are written as
\begin{eqnarray}
  {\cal L}_5 &=& \frac{1}{M_C}\epsilon_{\hat{a}\hat{b}\hat{c}}
  \left\{
      C(\wt{u}\wt{d}ul_L)^{MNij}
      \wt{u}^{\hat{a}}_M
      \wt{d}^{\hat{b}}_N
      ( u_{Li}^{\hat{c}}l_{Lj} )
    + C(\wt{u}\wt{u}dl_L)^{MNij} \frac{1}{2}
      \wt{u}^{\hat{a}}_M
      \wt{u}^{\hat{b}}_N
      ( d_{Li}^{\hat{c}}l_{Lj} )
  \right.
\nonumber\\&&\phantom{\frac{1}{M_C}\epsilon_{\hat{a}\hat{b}\hat{c}}}
  \left.
   +  C(\wt{u}\wt{d}ul_R)^{MNij}
      \wt{u}^{\hat{a}}_M
      \wt{d}^{\hat{b}}_N
      ( u_{Ri}^{\hat{c}}l_{Rj} )
    + C(\wt{u}\wt{u}dl_R)^{MNij} \frac{1}{2}
      \wt{u}^{\hat{a}}_M
      \wt{u}^{\hat{b}}_N
      ( d_{Ri}^{\hat{c}}l_{Rj} )
  \right.
\nonumber\\&&\phantom{\frac{1}{M_C}\epsilon_{\hat{a}\hat{b}\hat{c}}}
  \left.
    + C(\wt{u}\wt{d}d\nu_L)^{MNij}
      \wt{u}^{\hat{a}}_M
      \wt{d}^{\hat{b}}_N
      ( d_{Li}^{\hat{c}}\nu_{Lj} )
    + C(\wt{d}\wt{d}u\nu_L)^{MNij} \frac{1}{2}
      \wt{d}^{\hat{a}}_M
      \wt{d}^{\hat{b}}_N
      ( u_{Li}^{\hat{c}}\nu_{Lj} )
  \right.
\nonumber\\&&\phantom{\frac{1}{M_C}\epsilon_{\hat{a}\hat{b}\hat{c}}}
  \left.
    + C(\wt{u}\wt{l}ud_L)^{IJkl}
      \wt{u}^{\hat{a}}_I
      \wt{l}^{       }_J
      ( u_{Lk}^{\hat{b}}d_{Ll}^{\hat{c}} )
    + C(\wt{d}\wt{l}uu_L)^{IJkl} \frac{1}{2}
      \wt{d}^{\hat{a}}_I
      \wt{l}^{       }_J
      ( u_{Lk}^{\hat{b}}u_{Ll}^{\hat{c}} )
  \right.
\nonumber\\&&\phantom{\frac{1}{M_C}\epsilon_{\hat{a}\hat{b}\hat{c}}}
  \left.
   +  C(\wt{u}\wt{l}ud_R)^{IJkl}
      \wt{u}^{\hat{a}}_I
      \wt{l}^{       }_J
      ( u_{Rk}^{\hat{b}}d_{Rl}^{\hat{c}} )
   +  C(\wt{d}\wt{l}uu_R)^{IJkl} \frac{1}{2}
      \wt{d}^{\hat{a}}_I
      \wt{l}^{       }_J
      ( u_{Rk}^{\hat{b}}u_{Rl}^{\hat{c}} )
  \right.
\nonumber\\&&\phantom{\frac{1}{M_C}\epsilon_{\hat{a}\hat{b}\hat{c}}}
  \left.
   +  C(\wt{d}\wt{\nu}ud_L)^{Ijkl}
      \wt{d}^{\hat{a}}_I
      \wt{\nu}^{     }_j
      ( u_{Lk}^{\hat{b}}d_{Ll}^{\hat{c}} )
   +  C(\wt{u}\wt{\nu}dd_L)^{Ijkl} \frac{1}{2}
      \wt{u}^{\hat{a}}_I
      \wt{\nu}^{     }_j
      ( d_{Lk}^{\hat{b}}d_{Ll}^{\hat{c}} )
  \right\} ~,
\label{eq:dim5mass}
\end{eqnarray}
where the suffices $L,R$ of the quark/lepton fields denote the
chirality.
The coefficients $C$'s are written in terms of $C_{5L,R}$ as follows.
\begin{subeqnarray}
  C(\wt{u}\wt{d}ul_L)^{MNij} &=&
  \left( C_{5L}^{ijkl} - C_{5L}^{kjil} \right)
  \left( \wt{U}_U^\dagger \right)_{k}^{~M}
  \left( \wt{U}_D^\dagger \right)_{l}^{~N}
~,\\  
  C(\wt{u}\wt{u}dl_L)^{MNij} &=&
  \left( C_{5L}^{kjlm} - C_{5L}^{ljkm} \right)
  \left( \wt{U}_U^\dagger \right)_{k}^{~M}
  \left( \wt{U}_U^\dagger \right)_{l}^{~N}
  \left( V_{\rm KM} \right)_m^{~i}
~,\\  
  C(\wt{u}\wt{d}ul_R)^{MNij} &=&
  \left( C_{5R}^{*klji} - C_{5R}^{*iljk} \right)
  \left( \wt{U}_U^\dagger \right)_{k+3}^{~M}
  \left( \wt{U}_D^\dagger \right)_{l+3}^{~N}
~,\\  
  C(\wt{u}\wt{u}dl_R)^{MNij} &=&
  \left( C_{5R}^{*lijk} - C_{5R}^{*kijl} \right)
  \left( \wt{U}_U^\dagger \right)_{k+3}^{~M}
  \left( \wt{U}_U^\dagger \right)_{l+3}^{~N}
~,\\  
  C(\wt{u}\wt{d}d\nu_L)^{MNij} &=&
  \left( C_{5L}^{mjkl} - C_{5L}^{ljkm} \right)
  \left( \wt{U}_U^\dagger \right)_{k}^{~M}
  \left( \wt{U}_D^\dagger \right)_{l}^{~N}
  \left( V_{\rm KM} \right)_m^{~i}
~,\\  
  C(\wt{d}\wt{d}u\nu_L)^{MNij} &=&
  \left( C_{5L}^{ljik} - C_{5L}^{kjil} \right)
  \left( \wt{U}_D^\dagger \right)_{k}^{~M}
  \left( \wt{U}_D^\dagger \right)_{l}^{~N}
~,\\  
  C(\wt{u}\wt{l}ud_L)^{IJkl} &=&
  \left( C_{5L}^{ijkm} - C_{5L}^{kjim} \right)
  \left( \wt{U}_U^\dagger \right)_{i}^{~I}
  \left( \wt{U}_L         \right)_{j}^{~J}
  \left( V_{\rm KM} \right)_m^{~l}
~,\\  
  C(\wt{d}\wt{l}uu_L)^{IJkl} &=&
  \left( C_{5L}^{kjli} - C_{5L}^{ljki} \right)
  \left( \wt{U}_D^\dagger \right)_{i}^{~I}
  \left( \wt{U}_L         \right)_{j}^{~J}
~,\\  
  C(\wt{u}\wt{l}ud_R)^{IJkl} &=&
  \left( C_{5R}^{*klji} - C_{5R}^{*iljk} \right)
  \left( \wt{U}_U^\dagger \right)_{i+3}^{~I}
  \left( \wt{U}_L         \right)_{j+3}^{~J}
~,\\  
  C(\wt{d}\wt{l}uu_R)^{IJkl} &=&
  \left( C_{5R}^{*lijk} - C_{5R}^{*kijl} \right)
  \left( \wt{U}_D^\dagger \right)_{i+3}^{~I}
  \left( \wt{U}_L         \right)_{j+3}^{~J}
~,\\  
  C(\wt{d}\wt{\nu}ud_L)^{Ijkl} &=&
  \left( C_{5L}^{inkm} - C_{5L}^{mnki} \right)
  \left( \wt{U}_D^\dagger \right)_{i}^{~I}
  \left( \wt{U}_N         \right)_{n}^{~j}
  \left( V_{\rm KM} \right)_m^{~l}
~,\\  
  C(\wt{u}\wt{\nu}dd_L)^{Ijkl} &=&
  \left( C_{5L}^{qnip} - C_{5L}^{pniq} \right)
  \left( \wt{U}_U^\dagger \right)_{i}^{~I}
  \left( \wt{U}_N         \right)_{n}^{~j}
  \left( V_{\rm KM} \right)_p^{~k}
  \left( V_{\rm KM} \right)_q^{~l}
~.
\end{subeqnarray}
$C_{5L}$ and $C_{5R}$ at the electroweak scale are evaluated by solving
the renormalization group equations 
\begin{subeqnarray}
  \Dt C_{5L}^{ijkl} &=& 
  \left(
    -8g_3^2 -6g_2^2 -\frac{2}{3}g_1^2
  \right)C_{5L}^{ijkl}
\nonumber\\&&
   + C_{5L}^{mjkl}
     \left( f_D f_D^\dagger + f_U f_U^\dagger \right)^{i}_{~m}
   + C_{5L}^{imkl}
     \left( f_L^\dagger f_L \right)_{m}^{~j}
\nonumber\\&&
   + C_{5L}^{ijml}
     \left( f_D f_D^\dagger + f_U f_U^\dagger \right)^{k}_{~m}
   + C_{5L}^{ijkm}
     \left( f_D f_D^\dagger + f_U f_U^\dagger \right)^{l}_{~m} ~,
\\
  \Dt C_{5R}^{ijkl} &=& 
  \left(
    -8g_3^2 -4g_1^2
  \right)C_{5R}^{ijkl}
\nonumber\\&&
   + C_{5R}^{mjkl}
     \left( 2\, f_U^\dagger f_U \right)_{m}^{~i}
   + C_{5R}^{imkl}
     \left( 2\, f_D^\dagger f_D \right)_{m}^{~j}
\nonumber\\&&
   + C_{5R}^{ijml}
     \left( 2\, f_L f_L^\dagger \right)^{k}_{~m}
   + C_{5R}^{ijkm}
     \left( 2\, f_U^\dagger f_U \right)_{m}^{~l} ~,
\end{subeqnarray}
where $\Lambda$ is the renormalization point.

\subsubsection{Effective interactions}   

After the calculation of the one-loop (gluino-, chargino- and
neutralino-) dressing diagrams, effective four-fermi interaction terms
relevant to the nucleon decay are obtained as follows.
\begin{eqnarray}
  {\cal L}_{\not{B}} &=& \frac{1}{(4\pi)^2 M_C}
  \epsilon_{\hat{a}\hat{b}\hat{c}}
  \left\{
      \wt{C}_{LL}(udul)^{ik}
      (u_L^{\hat{a}}d_{Li}^{\hat{b}})(u_L^{\hat{c}}l_{Lk})
   +  \wt{C}_{RL}(udul)^{ik}
      (u_R^{\hat{a}}d_{Ri}^{\hat{b}})(u_L^{\hat{c}}l_{Lk})
  \right.
  \nonumber\\&&\phantom{\epsilon_{\hat{a}\hat{b}\hat{c}}}
  \left.
   +  \wt{C}_{LR}(udul)^{ik}
      (u_L^{\hat{a}}d_{Li}^{\hat{b}})(u_R^{\hat{c}}l_{Rk})
   +  \wt{C}_{RR}(udul)^{ik}
      (u_R^{\hat{a}}d_{Ri}^{\hat{b}})(u_R^{\hat{c}}l_{Rk})
  \right.
  \nonumber\\&&\phantom{\epsilon_{\hat{a}\hat{b}\hat{c}}}
  \left.
   +  \wt{C}_{LL}(udd\nu)^{ijk}
      (u_L^{\hat{a}}d_{Li}^{\hat{b}})(d_{Lj}^{\hat{c}}\nu_{Lk})
   +  \wt{C}_{RL}(udd\nu)^{ijk}
      (u_R^{\hat{a}}d_{Ri}^{\hat{b}})(d_{Lj}^{\hat{c}}\nu_{Lk})
  \right.
  \nonumber\\&&\phantom{\epsilon_{\hat{a}\hat{b}\hat{c}}}
  \left.
   +  \wt{C}_{RL}(ddu\nu)^{ijk}\frac{1}{2}
      (d_{Ri}^{\hat{a}}d_{Rj}^{\hat{b}})(u_L^{\hat{c}}\nu_{Lk})
  \right\} ~,
\label{eq:lag3q}
\end{eqnarray}
\begin{subeqnarray}
  \wt{C}_{LL}(udul)^{ik} &=&
        \wt{C}_{LL}(udul)^{ik}_{\wt{G}}
      + \wt{C}_{LL}(udul)^{ik}_{\chi^{\pm}}
      + \wt{C}_{LL}(udul)^{ik}_{\chi^0}
~,
\\
  \wt{C}_{LL}(udul)^{ik}_{\wt{G}} &=&
  \frac{4}{3}\frac{g_3^2}{M_{\wt{G}}}
  C(\wt{u}\wt{d}ul_L)^{MN1k}
  \left( \Gamma_{GL}^{(u)} \right)_{M}^{1}
  \left( \Gamma_{GL}^{(d)} \right)_{N}^{i}
  H( u_M^{\wt{G}}, x_N^{\wt{G}} )
~,
\\
  \wt{C}_{LL}(udul)^{ik}_{\chi^{\pm}} &=&
  \frac{g_2^2}{M_{C}^\alpha}
  \left[
    - C(\wt{u}\wt{d}ul_L)^{MN1k}
    \left( \Gamma_{CL}^{(u)} \right)_{N}^{\alpha 1}
    \left( \Gamma_{CL}^{(d)} \right)_{M}^{\alpha i}
    H( x_M^{\alpha}, u_N^{\alpha} )
  \right.
  \nonumber\\&&\phantom{
  \frac{g_2^2}{M_{C}^\alpha}}
  \left.
    + C(\wt{d}\wt{\nu}ud_L)^{Nm1i}
    \left( \Gamma_{CL}^{(u)} \right)_{N}^{\alpha 1}
    \left( \Gamma_{CL}^{(l)} \right)_{m}^{\alpha k}
    H( u_N^{\alpha}, z_m^{\alpha} )
  \right]
~,\\
  \wt{C}_{LL}(udul)^{ik}_{\chi^0} &=&
  \frac{g_2^2}{M_{N}^{\ol{\alpha}}}
  \left[
    C(\wt{u}\wt{d}ul_L)^{MN1k}
    \left( \Gamma_{NL}^{(u)} \right)_{M}^{\ol{\alpha} 1}
    \left( \Gamma_{NL}^{(d)} \right)_{N}^{\ol{\alpha} i}
    H( v_M^{\ol{\alpha}}, y_N^{\ol{\alpha}} )
  \right.
  \nonumber\\&&\phantom{
  \frac{g_2^2}{M_{N}^{\ol{\alpha}}}}
  \left.
    + C(\wt{u}\wt{l}ud_L)^{MN1i}
    \left( \Gamma_{NL}^{(u)} \right)_{M}^{\ol{\alpha} 1}
    \left( \Gamma_{NL}^{(l)} \right)_{N}^{\ol{\alpha} k}
    H( v_M^{\ol{\alpha}}, z_N^{\ol{\alpha}} )
  \right]
~,
\end{subeqnarray}
\begin{subeqnarray}
  \wt{C}_{RL}(udul)^{ik} &=&
  \wt{C}_{RL}^{(6)}(udul)^{ik}
  \nonumber\\&&    
      + \wt{C}_{RL}(udul)^{ik}_{\wt{G}}
      + \wt{C}_{RL}(udul)^{ik}_{\chi^{\pm}}
      + \wt{C}_{RL}(udul)^{ik}_{\chi^0}
~,\\
  \wt{C}_{RL}(udul)^{ik}_{\wt{G}} &=&
  \frac{4}{3}\frac{g_3^2}{M_{\wt{G}}}
  C(\wt{u}\wt{d}ul_L)^{MN1k}
  \left( \Gamma_{GR}^{(u)} \right)_{M}^{1}
  \left( \Gamma_{GR}^{(d)} \right)_{N}^{i}
  H( u_M^{\wt{G}}, x_N^{\wt{G}} )
~,
\\
  \wt{C}_{RL}(udul)^{ik}_{\chi^{\pm}} &=&
  - \frac{g_2^2}{M_{C}^\alpha}
    C(\wt{u}\wt{d}ul_L)^{MN1k}
    \left( \Gamma_{CR}^{(u)} \right)_{N}^{\alpha 1}
    \left( \Gamma_{CR}^{(d)} \right)_{M}^{\alpha i}
    H( x_M^{\alpha}, u_N^{\alpha} )
~,
\\
  \wt{C}_{RL}(udul)^{ik}_{\chi^0} &=&
  \frac{g_2^2}{M_{N}^{\ol{\alpha}}}
  \left[
    C(\wt{u}\wt{d}ul_L)^{MN1k}
    \left( \Gamma_{NR}^{(u)} \right)_{M}^{\ol{\alpha} 1}
    \left( \Gamma_{NR}^{(d)} \right)_{N}^{\ol{\alpha} i}
    H( v_M^{\ol{\alpha}}, y_N^{\ol{\alpha}} )
  \right.
  \nonumber\\&&\phantom{
  \frac{g_2^2}{M_{N}^{\ol{\alpha}}}}
  \left.
    + C(\wt{u}\wt{l}ud_R)^{MN1i}
    \left( \Gamma_{NL}^{(u)} \right)_{M}^{\ol{\alpha} 1}
    \left( \Gamma_{NL}^{(l)} \right)_{N}^{\ol{\alpha} k}
    H( v_M^{\ol{\alpha}}, z_N^{\ol{\alpha}} )
  \right]
~,
\end{subeqnarray}
\begin{subeqnarray}
  \wt{C}_{LR}(udul)^{ik} &=&
  \wt{C}^{(6)}_{LR}(udul)^{ik}
  \nonumber\\&&    
      + \wt{C}_{LL}(udul)^{ik}_{\wt{G}}
      + \wt{C}_{LL}(udul)^{ik}_{\chi^{\pm}}
      + \wt{C}_{LL}(udul)^{ik}_{\chi^0}
~,\\
  \wt{C}_{LR}(udul)^{ik}_{\wt{G}} &=&
  \frac{4}{3}\frac{g_3^2}{M_{\wt{G}}}
  C(\wt{u}\wt{d}ul_R)^{MN1k}
  \left( \Gamma_{GL}^{(u)} \right)_{M}^{1}
  \left( \Gamma_{GL}^{(d)} \right)_{N}^{i}
  H( u_M^{\wt{G}}, x_N^{\wt{G}} )
~,\\
  \wt{C}_{LR}(udul)^{ik}_{\chi^{\pm}} &=&
  \frac{g_2^2}{M_{C}^\alpha}
  \left[
    - C(\wt{u}\wt{d}ul_R)^{MN1k}
    \left( \Gamma_{CL}^{(u)} \right)_{N}^{\alpha 1}
    \left( \Gamma_{CL}^{(d)} \right)_{M}^{\alpha i}
    H( x_M^{\alpha}, u_N^{\alpha} )
  \right.
  \nonumber\\&&\phantom{
  \frac{g_2^2}{M_{C}^\alpha}}
  \left.
    + C(\wt{d}\wt{\nu}ud_L)^{Nm1i}
    \left( \Gamma_{CR}^{(u)} \right)_{N}^{\alpha 1}
    \left( \Gamma_{CR}^{(l)} \right)_{m}^{\alpha k}
    H( u_N^{\alpha}, z_m^{\alpha} )
  \right]
~,\\
  \wt{C}_{LR}(udul)^{ik}_{\chi^0} &=&
  \frac{g_2^2}{M_{N}^{\ol{\alpha}}}
  \left[
    C(\wt{u}\wt{d}ul_R)^{MN1k}
    \left( \Gamma_{NL}^{(u)} \right)_{M}^{\ol{\alpha} 1}
    \left( \Gamma_{NL}^{(d)} \right)_{N}^{\ol{\alpha} i}
    H( v_M^{\ol{\alpha}}, y_N^{\ol{\alpha}} )
  \right.
  \nonumber\\&&\phantom{
  \frac{g_2^2}{M_{N}^{\ol{\alpha}}}}
  \left.
    + C(\wt{u}\wt{l}ud_L)^{MN1i}
    \left( \Gamma_{NR}^{(u)} \right)_{M}^{\ol{\alpha} 1}
    \left( \Gamma_{NR}^{(l)} \right)_{N}^{\ol{\alpha} k}
    H( v_M^{\ol{\alpha}}, z_N^{\ol{\alpha}} )
  \right]
~,
\end{subeqnarray}
\begin{subeqnarray}
  \wt{C}_{RR}(udul)^{ik} &=&
        \wt{C}_{RR}(udul)^{ik}_{\wt{G}}
      + \wt{C}_{RR}(udul)^{ik}_{\chi^{\pm}}
      + \wt{C}_{RR}(udul)^{ik}_{\chi^0}
~,
\\
  \wt{C}_{RR}(udul)^{ik}_{\wt{G}} &=&
  \frac{4}{3}\frac{g_3^2}{M_{\wt{G}}}
  C(\wt{u}\wt{d}ul_R)^{MN1k}
  \left( \Gamma_{GR}^{(u)} \right)_{M}^{1}
  \left( \Gamma_{GR}^{(d)} \right)_{N}^{i}
  H( u_M^{\wt{G}}, x_N^{\wt{G}} )
~,\\
  \wt{C}_{RR}(udul)^{ik}_{\chi^{\pm}} &=&
  - \frac{g_2^2}{M_{C}^\alpha}
    C(\wt{u}\wt{d}ul_R)^{MN1k}
    \left( \Gamma_{CR}^{(u)} \right)_{N}^{\alpha 1}
    \left( \Gamma_{CR}^{(d)} \right)_{M}^{\alpha i}
    H( x_M^{\alpha}, u_N^{\alpha} )
~,\\
  \wt{C}_{RR}(udul)^{ik}_{\chi^0} &=&
  \frac{g_2^2}{M_{N}^{\ol{\alpha}}}
  \left[
    C(\wt{u}\wt{d}ul_R)^{MN1k}
    \left( \Gamma_{NR}^{(u)} \right)_{M}^{\ol{\alpha} 1}
    \left( \Gamma_{NR}^{(d)} \right)_{N}^{\ol{\alpha} i}
    H( v_M^{\ol{\alpha}}, y_N^{\ol{\alpha}} )
  \right.
  \nonumber\\&&\phantom{
  \frac{g_2^2}{M_{N}^{\ol{\alpha}}}}
  \left.
    + C(\wt{u}\wt{l}ud_R)^{MN1i}
    \left( \Gamma_{NR}^{(u)} \right)_{M}^{\ol{\alpha} 1}
    \left( \Gamma_{NR}^{(l)} \right)_{N}^{\ol{\alpha} k}
    H( v_M^{\ol{\alpha}}, z_N^{\ol{\alpha}} )
  \right]
~,
\end{subeqnarray}
\begin{subeqnarray}
  \wt{C}_{LL}(udd\nu)^{ijk} &=&
        \wt{C}_{LL}(udd\nu)^{ijk}_{\wt{G}}
      + \wt{C}_{LL}(udd\nu)^{ijk}_{\chi^{\pm}}
      + \wt{C}_{LL}(udd\nu)^{ijk}_{\chi^0}
~,
\\
  \wt{C}_{LL}(udd\nu)^{ijk}_{\wt{G}} &=&
  \frac{4}{3}\frac{g_3^2}{M_{\wt{G}}}
  \left[
    C(\wt{u}\wt{d}d\nu_L)^{MNjk}
    \left( \Gamma_{GL}^{(u)} \right)_{M}^{1}
    \left( \Gamma_{GL}^{(d)} \right)_{N}^{i}
    H( u_M^{\wt{G}}, x_N^{\wt{G}} )
  \right.
  \nonumber\\&&\phantom{\frac{4}{3}\frac{g_3^2}{M_{\wt{G}}}}
  \left.
  + C(\wt{d}\wt{d}u\nu_L)^{MN1k}
    \left( \Gamma_{GL}^{(d)} \right)_{M}^{j}
    \left( \Gamma_{GL}^{(d)} \right)_{N}^{i}
    H( x_M^{\wt{G}}, x_N^{\wt{G}} )
  \right]
~,\\
  \wt{C}_{LL}(udd\nu)^{ijk}_{\chi^{\pm}} &=&
  \frac{g_2^2}{M_{C}^\alpha}
  \left[
    - C(\wt{u}\wt{d}d\nu_L)^{MNjk}
    \left( \Gamma_{CL}^{(u)} \right)_{N}^{\alpha 1}
    \left( \Gamma_{CL}^{(d)} \right)_{M}^{\alpha i}
    H( x_M^{\alpha}, u_N^{\alpha} )
  \right.
  \nonumber\\&&\phantom{
  \frac{g_2^2}{M_{C}^\alpha}}
  \left.
    + C(\wt{u}\wt{l}ud_L)^{MN1i}
    \left( \Gamma_{CL}^{(d)} \right)_{M}^{\alpha j}
    \left( \Gamma_{CL}^{(\nu)} \right)_{N}^{\alpha k}
    H( x_M^{\alpha}, w_N^{\alpha} )
  \right]
~,\\
  \wt{C}_{LL}(udd\nu)^{ijk}_{\chi^0} &=&
  \frac{g_2^2}{M_{N}^{\ol{\alpha}}}
  \left[
    C(\wt{u}\wt{d}d\nu_L)^{MNjk}
    \left( \Gamma_{NL}^{(u)} \right)_{M}^{\ol{\alpha} 1}
    \left( \Gamma_{NL}^{(d)} \right)_{N}^{\ol{\alpha} i}
    H( v_M^{\ol{\alpha}}, y_N^{\ol{\alpha}} )
  \right.
  \nonumber\\&&\phantom{
  \frac{g_2^2}{M_{N}^{\ol{\alpha}}}}
  \left.
    + C(\wt{d}\wt{d}u\nu_L)^{MN1k}
    \left( \Gamma_{NL}^{(d)} \right)_{M}^{\ol{\alpha} j}
    \left( \Gamma_{NL}^{(d)} \right)_{N}^{\ol{\alpha} i}
    H( y_M^{\ol{\alpha}}, y_N^{\ol{\alpha}} )
  \right.
  \nonumber\\&&\phantom{
  \frac{g_2^2}{M_{N}^{\ol{\alpha}}}}
  \left.
    + C(\wt{d}\wt{\nu}ud_L)^{Mn1i}
    \left( \Gamma_{NL}^{(d)} \right)_{M}^{\ol{\alpha} j}
    \left( \Gamma_{NL}^{(\nu)} \right)_{n}^{\ol{\alpha} k}
    H( y_M^{\ol{\alpha}}, w_n^{\ol{\alpha}} )
  \right.
  \nonumber\\&&\phantom{
  \frac{g_2^2}{M_{N}^{\ol{\alpha}}}}
  \left.
    + C(\wt{u}\wt{\nu}dd_L)^{Mnji}
    \left( \Gamma_{NL}^{(u)} \right)_{M}^{\ol{\alpha} 1}
    \left( \Gamma_{NL}^{(\nu)} \right)_{n}^{\ol{\alpha} k}
    H( v_M^{\ol{\alpha}}, w_n^{\ol{\alpha}} )
  \right]
~,
\end{subeqnarray}
\begin{subeqnarray}
  \wt{C}_{RL}(udd\nu)^{ijk} &=&
  \wt{C}^{(6)}_{RL}(udd\nu)^{ijk}
  \nonumber\\&&    
      + \wt{C}_{RL}(udd\nu)^{ijk}_{\wt{G}}
      + \wt{C}_{RL}(udd\nu)^{ijk}_{\chi^{\pm}}
      + \wt{C}_{RL}(udd\nu)^{ijk}_{\chi^0}
~,
\\
  \wt{C}_{RL}(udd\nu)^{ijk}_{\wt{G}} &=&
  \frac{4}{3}\frac{g_3^2}{M_{\wt{G}}}
    C(\wt{u}\wt{d}d\nu_L)^{MNjk}
    \left( \Gamma_{GR}^{(u)} \right)_{M}^{1}
    \left( \Gamma_{GR}^{(d)} \right)_{N}^{i}
    H( u_M^{\wt{G}}, x_N^{\wt{G}} )
~,
\\
  \wt{C}_{RL}(udd\nu)^{ijk}_{\chi^{\pm}} &=&
  \frac{g_2^2}{M_{C}^\alpha}
  \left[
    - C(\wt{u}\wt{d}d\nu_L)^{MNjk}
    \left( \Gamma_{CR}^{(u)} \right)_{N}^{\alpha 1}
    \left( \Gamma_{CR}^{(d)} \right)_{M}^{\alpha i}
    H( x_M^{\alpha}, u_N^{\alpha} )
  \right.
  \nonumber\\&&\phantom{
  \frac{g_2^2}{M_{C}^\alpha}}
  \left.
    + C(\wt{u}\wt{l}ud_R)^{MN1i}
    \left( \Gamma_{CL}^{(d)} \right)_{M}^{\alpha j}
    \left( \Gamma_{CL}^{(\nu)} \right)_{N}^{\alpha k}
    H( x_M^{\alpha}, w_N^{\alpha} )
  \right]
~,
\\
  \wt{C}_{RL}(udd\nu)^{ijk}_{\chi^0} &=&
  \frac{g_2^2}{M_{N}^{\ol{\alpha}}}
    C(\wt{u}\wt{d}d\nu_L)^{MNjk}
    \left( \Gamma_{NR}^{(u)} \right)_{M}^{\ol{\alpha} 1}
    \left( \Gamma_{NR}^{(d)} \right)_{N}^{\ol{\alpha} i}
    H( v_M^{\ol{\alpha}}, y_N^{\ol{\alpha}} )
~,
\end{subeqnarray}
\begin{subeqnarray}
  \wt{C}_{RL}(ddu\nu)^{ijk} &=&
      \wt{C}_{RL}(ddu\nu)^{ijk}_{\wt{G}}
    + \wt{C}_{RL}(ddu\nu)^{ijk}_{\chi^0}
~,
\\
  \wt{C}_{RL}(ddu\nu)^{ijk}_{\wt{G}} &=&
  \frac{4}{3}\frac{g_3^2}{M_{\wt{G}}}
    C(\wt{d}\wt{d}u\nu_L)^{MN1k}
    \left( \Gamma_{GR}^{(d)} \right)_{M}^{i}
    \left( \Gamma_{GR}^{(d)} \right)_{N}^{j}
    H( x_M^{\wt{G}}, x_N^{\wt{G}} )
~,
\\
  \wt{C}_{RL}(ddu\nu)^{ijk}_{\chi^0} &=&
  \frac{g_2^2}{M_{N}^{\ol{\alpha}}}
    C(\wt{d}\wt{d}u\nu_L)^{MN1k}
    \left( \Gamma_{NR}^{(d)} \right)_{M}^{\ol{\alpha} i}
    \left( \Gamma_{NR}^{(d)} \right)_{N}^{\ol{\alpha} j}
    H( y_M^{\ol{\alpha}}, y_N^{\ol{\alpha}} )
~.
\end{subeqnarray}
Here, $\wt{C}_{RL,LR}^{(6)}$ are contributions from dimension six
operators, whose magnitudes are quite small compared to the dimension
five contributions for $B \rightarrow M \ol{\nu}$ decay modes
($B=p$ or $n$, $M=K$, $\pi$ or $\eta$).
Notice that $C(\wt{u}\wt{u}dl_{L,R})$ and  $C(\wt{d}\wt{l}uu_{L,R})$ in
(\ref{eq:dim5mass}) do not contribute to the nucleon decay amplitude.
The function $H$ is defined as
\begin{eqnarray}
  H(x,y) &=&
  \frac{1}{x-y}
  \left(
      \frac{x\log x}{x-1} - \frac{y\log y}{y-1}
  \right) ~,
\end{eqnarray}
and the arguments of $H$ are ratios of SUSY particles' masses (squared):
\begin{subeqnarray}
  x_{M}^{\wt{G}} &=&
  \frac{ m_{\wt{d}_M}^2 }{ M_{\wt{G}}^2 } ~,
~~~
  u_{M}^{\wt{G}} ~=~
  \frac{ m_{\wt{u}_M}^2 }{ M_{\wt{G}}^2 } ~,
\\
  x_{M}^{\alpha} &=&
  \frac{ m_{\wt{u}_M}^2 }{ M_{C}^{\alpha 2} } ~,
~~~
  u_{M}^{\alpha} ~=~
  \frac{ m_{\wt{d}_M}^2 }{ M_{C}^{\alpha 2} } ~,
~~~
  z_{m}^{\alpha} ~=~
  \frac{ m_{\wt{\nu}_m}^2 }{ M_{C}^{\alpha 2} } ~,
~~~
  w_{M}^{\alpha} ~=~
  \frac{ m_{\wt{l}_M}^2 }{ M_{C}^{\alpha 2} } ~,
\\
  v_{M}^{\ol{\alpha}} &=&
  \frac{ m_{\wt{u}_M}^2 }{ M_{N}^{\ol{\alpha} 2} }~,
~~~
  y_{M}^{\ol{\alpha}} ~=~
  \frac{ m_{\wt{d}_M}^2 }{ M_{N}^{\ol{\alpha} 2} } ~,
~~~
  z_{M}^{\ol{\alpha}} ~=~
  \frac{ m_{\wt{l}_M}^2 }{ M_{N}^{\ol{\alpha} 2} } ~,
~~~
  w_{m}^{\ol{\alpha}} ~=~
  \frac{ m_{\wt{\nu}_m}^2 }{ M_{N}^{\ol{\alpha} 2} } ~.
\end{subeqnarray}

\subsubsection{Nucleon partial decay widths}

The effective quark Lagrangian (\ref{eq:lag3q}) is converted to an
effective hadronic Lagrangian with use of the chiral Lagrangian
technique (perturbative QCD corrections between the electroweak
scale and $\sim 1$ GeV scale are also taken into account), then partial
decay widths of the nucleon are calculated as
\begin{eqnarray}
  \Gamma( B_i \rightarrow M_j l_k ) &=&
  \frac{m_i}{32\pi}\left( 1 - \frac{m_j^2}{m_i^2} \right)^2
  \frac{1}{f_\pi^2}
  \left(
    \left| A_L^{ijk} \right|^2
  + \left| A_R^{ijk} \right|^2
  \right)
~,
\label{eq:pdecayrate}
\end{eqnarray}
where the lepton mass is neglected only for the kinematics.
The expressions for $A_{L,R}^{ijk}$ are listed in
Table~\ref{tab:pdecayamp}.

%

%
\newpage
%
%
%
%
%
%
%
%
%
%


\begin{table}
\begin{center}
%

\renewcommand{\arraystrut}{%
\protect\rule[-1.6ex]{0em}{5.2ex}\protect\rule{\arraycolsep}{0ex}%
}
\begin{displaymath}
\makebox[0em]{$
\begin{array}{@{\arraystrut}c|c|c|c|l}
\multicolumn{5}{c}{\arraystrut}\\
\hline
\hline
 B_i & l_k & M_j & & \multicolumn{1}{c}{A_L^{ijk}~,\ A_R^{ijk} } \\
\hline
 p & l^+_k & \pi^0 & L &
 \phantom{+}\frac{1}{\sqrt{2}}(1+F+D)
 \left[
   \alpha_p \wt{C}_{RL}(udul)^{1k}
 + \beta_p  \wt{C}_{LL}(udul)^{1k}
 \right]  \\
   &  &  & R &
 -\frac{1}{\sqrt{2}}(1+F+D)
 \left[
   \alpha_p \wt{C}_{LR}(udul)^{1k}
 + \beta_p  \wt{C}_{RR}(udul)^{1k}
 \right]  \\
\cline{3-5}
   &       & \eta^0 & L &
  \phantom{+}\sqrt{\frac{3}{2}} (-\frac{1}{3}+F-\frac{1}{3}D)\,
   \alpha_p \wt{C}_{RL}(udul)^{1k}
 +\sqrt{\frac{3}{2}} (1+F-\frac{1}{3}D)\,
   \beta_p  \wt{C}_{LL}(udul)^{1k}
\\
   &       &        & R &
  -\sqrt{\frac{3}{2}} (-\frac{1}{3}+F-\frac{1}{3}D)\,
   \alpha_p \wt{C}_{LR}(udul)^{1k}
 -\sqrt{\frac{3}{2}} (1+F-\frac{1}{3}D)\,
   \beta_p  \wt{C}_{RR}(udul)^{1k}
\\
\cline{3-5}
   &       & K^0 & L &
 \phantom{+}\left(
   -1 + \frac{m_N}{m_{B'}}(F-D)
 \right)
 \alpha_p \wt{C}_{RL}(udul)^{2k}
 +\left(
   1 + \frac{m_N}{m_{B'}}(F-D)
 \right)
 \beta_p \wt{C}_{LL}(udul)^{2k}
\\
   &       &  & R &
 -\left(
   -1 + \frac{m_N}{m_{B'}}(F-D)
 \right)
 \alpha_p \wt{C}_{LR}(udul)^{2k}
 -\left(
   1 + \frac{m_N}{m_{B'}}(F-D)
 \right)
 \beta_p \wt{C}_{RR}(udul)^{2k}
\\
\cline{2-5}
   & \ol{\nu}_k & \pi^+ & L &
 (1+F+D)
 \left[
   \alpha_p \wt{C}_{RL}(udd\nu)^{11k}
 + \beta_p  \wt{C}_{LL}(udd\nu)^{11k}
 \right]
\\
\cline{3-5}
   &            & K^+ & L &
 \left(
   1 - \frac{m_N}{m_{B'}}(F-\frac{1}{3}D)
 \right)
 \alpha_p \wt{C}_{RL}(ddu\nu)^{12k}
 +
 \left(
   1 + \frac{m_N}{m_{B'}}(F+\frac{1}{3}D)
 \right)
 \alpha_p \wt{C}_{RL}(udd\nu)^{12k}
\\
   &            &  &  &
 +
 \left(
   \frac{m_N}{m_{B'}}\frac{2}{3}D
 \right)
 \alpha_p \wt{C}_{RL}(udd\nu)^{21k}
 +
 \left(
   1 + \frac{m_N}{m_{B'}}(F+\frac{1}{3}D)
 \right)
 \beta_p \wt{C}_{LL}(udd\nu)^{12k}
\\
   &            &  &  &
 +
 \left(
   \frac{m_N}{m_{B'}}\frac{2}{3}D
 \right)
 \beta_p \wt{C}_{LL}(udd\nu)^{21k}
\\
\hline
 n & l^+_k & \pi^- & L &
 \phantom{+}(1+F+D)
 \left[
   \alpha_p \wt{C}_{RL}(udul)^{1k}
 + \beta_p  \wt{C}_{LL}(udul)^{1k}
 \right]
\\
   &  &  & R &
 -(1+F+D)
 \left[
   \alpha_p \wt{C}_{LR}(udul)^{1k}
 + \beta_p  \wt{C}_{RR}(udul)^{1k}
 \right]
\\
\cline{2-5}
   & \ol{\nu}_k & \pi^0 & L &
 -\frac{1}{\sqrt{2}}(1+F+D)
 \left[
   \alpha_p \wt{C}_{RL}(udd\nu)^{11k}
 + \beta_p  \wt{C}_{LL}(udd\nu)^{11k}
 \right]
\\
\cline{3-5}
   &  & \eta^0 & L &
 \sqrt{\frac{3}{2}}(-\frac{1}{3}+F-\frac{1}{3}D)\,
   \alpha_p \wt{C}_{RL}(udd\nu)^{11k}
 +\sqrt{\frac{3}{2}}(1+F-\frac{1}{3}D)\,
   \beta_p  \wt{C}_{LL}(udd\nu)^{11k}
\\
\cline{3-5}
   &  & K^0 & L &
 \left(
   -\frac{m_N}{m_{B'}}\frac{2}{3}D
 \right)
 \alpha_p\wt{C}_{RL}(ddu\nu)^{12k}
 +
 \left(
   1+\frac{m_N}{m_{B'}}(F+\frac{1}{3}D)
 \right)
 \alpha_p\wt{C}_{RL}(udd\nu)^{12k}
\\
   &  &  &  &
 +
 \left(
   -1+\frac{m_N}{m_{B'}}(F-\frac{1}{3}D)
 \right)
 \alpha_p\wt{C}_{RL}(udd\nu)^{21k}
 +
 \left(
   1+\frac{m_N}{m_{B'}}(F+\frac{1}{3}D)
 \right)
 \beta_p\wt{C}_{LL}(udd\nu)^{12k}
\\
   &  &  &  &
 +
 \left(
   1+\frac{m_N}{m_{B'}}(F-\frac{1}{3}D)
 \right)
 \beta_p\wt{C}_{LL}(udd\nu)^{21k}
\\
\hline
\hline
\multicolumn{5}{c}{\arraystrut}
\end{array}
$}
\end{displaymath}
%
\end{center}
\caption[tabI]{$A_{L,R}^{ijk}$ in (\protect\ref{eq:pdecayrate}) 
for each nucleon decay mode.
$m_{N}$ is the nucleon mass $m_{N} \approx m_p \approx m_n$ and $m_{B'}$
is an averaged baryon mass $m_{B'} \approx m_\Sigma \approx m_\Lambda$.
$F\approx 0.48$ and $D\approx 0.76$ are coupling constants for the
interaction between baryons and mesons
\protect\cite{RRRR,Chiral_Lagrangian}.}
\label{tab:pdecayamp}
\end{table}
%
%
%
%
%
%

\begin{thebibliography}{99}
%
%
\bibitem{Gauge_Coupling_Unification}
P. Langacker and M.-X. Luo, 
Phys. Rev. {\bf D44} (1991) 817; 
U. Amaldi, W. de Boer and H. F\"{u}rstenau, 
Phys. Lett. {\bf B260} (1991) 447; 
W. J. Marciano, 
Ann. Rev. Nucl. Part. {\bf 41} (1991) 469. 
%
%
\bibitem{SUSY_GUT}
E. Witten, 
Nucl. Phys. {\bf B188} (1981) 513; 
S. Dimopoulos, S. Raby and F. Wilczek, 
Phys. Rev. {\bf D24} (1981) 1681; 
S. Dimopoulos and H. Georgi,
Nucl. Phys. {\bf B193} (1981) 150; 
N. Sakai,
Z. Phys. {\bf C11} (1981) 153.
%
\bibitem{DRW}
S. Dimopoulos, S. Raby and F. Wilczek, 
Phys. Lett. {\bf 112B} (1982) 133. 
%
\bibitem{ENR}
J. Ellis, D.V. Nanopoulos and S. Rudaz, 
Nucl. Phys. {\bf B202} (1982) 43.
%
%
\bibitem{SUGRA}
For reviews on the minimal SU(5) SUGRA GUT model, see for instance, 
H.P. Nilles, 
Phys. Rep. {\bf 110} (1984) 1; 
P. Nath, R. Arnowitt and A.H. Chamseddine, 
Applied $N=1$ Supergravity (World Scientific, Singapore, 1984). 
%
\bibitem{Kam}
Kamiokande Collaboration, K.S. Hirata {\it et al.}, 
Phys. Lett. {\bf B220} (1989) 308. 
%
\bibitem{IMB}
IMB Collaboration, R. Becker-Szendy {\it et al.}, 
Proceedings of 23rd International 
Cosmic Ray Conference, Calgary 1993 {\bf Vol.4} 589.
%
\bibitem{superK}
M. Takita (Super-Kamiokande Collaboration),
Talk presented in 29th International Conference on High Energy 
Physics, Vancouver, July 1998.
%
%
%
\bibitem{dim5_op}
N. Sakai and T. Yanagida, 
Nucl. Phys. {\bf B197} (1982) 533; 
S. Weinberg, 
Phys. Rev. {\bf D26} (1982) 287.
\bibitem{NCA}
P. Nath, A.H. Chamseddine and R. Arnowitt, 
Phys. Rev. {\bf D32} (1985) 2348.
%
\bibitem{MATS+HMY}
M. Matsumoto, J. Arafune, H. Tanaka and K. Shiraishi, 
Phys. Rev. {\bf D46} (1992) 3966; 
J. Hisano, H. Murayama and T. Yanagida, 
Nucl. Phys. {\bf B402} (1993) 46.
%
\bibitem{HMTY}
J. Hisano, T. Moroi, K. Tobe and T. Yanagida, 
Mod. Phys. Lett. {\bf A10} (1995) 2267.
%
\bibitem{GNA}
T. Goto, T. Nihei and J. Arafune, 
Phys. Rev. {\bf D52} (1995) 505.
%
%
\bibitem{RRRR}
V. Lucas and S. Raby, 
Phys. Rev. {\bf D55} (1997) 6986.
%
\bibitem{Phase_Matrix}
J. Ellis, M.K. Gaillard and D.V. Nanopoulos, 
Phys. Lett. {\bf 88B} (1979) 320. 
%
%
%
\bibitem{BKS}
A. Bouquet, J. Kaplan, and C.A. Savoy, 
Phys. Lett. {\bf 148B} (1984) 69; 
Nucl. Phys. {\bf B262} (1985) 299.
%
%
%
\bibitem{Radiative_Breaking}
K. Inoue, A. Kakuto, H. Komatsu and S. Takeshita,
Prog. Theor. Phys. {\bf 68} (1982) 927;
{\it ibid.}\ {\bf 71} (1984) 413; 
L. Ib\'{a}\~{n}ez and G.G. Ross,
Phys. Lett. {\bf 110B} (1982) 215; 
L. Alvarez-Gaum\'{e}, J. Polchinski and M.B. Wise,
Nucl. Phys. {\bf B221} (1983) 495; 
J. Ellis, J.S. Hagelin, D.V. Nanopoulos and K. Tamvakis,
Phys. Lett. {\bf 125B} (1983) 275.
%
\bibitem{Chiral_Lagrangian}
M. Claudson, M.B. Wise and L.J. Hall,
Nucl. Phys. {\bf B195} (1982) 297; 
S. Chadha and M. Daniel, 
Nucl. Phys. {\bf B229} (1983) 105.
%
%
\bibitem{beta_p}
S.J. Brodsky, J. Ellis, J.S. Hagelin and C.T. Sachrajda,
Nucl. Phys. {\bf B238} (1984) 561; 
M.B. Gavela, S.F. King, C.T. Sachrajda, G. Martinelli,
M.L. Paciello and B. Taglienti, 
Nucl. Phys. {\bf B312} (1989) 269.
%
%
%
%
\bibitem{PDG}
Particle Data Group, C. Caso {\it et al}., 
Euro. Phys. Journ. {\bf C3} (1998) 1.
%
%
\bibitem{m_top}
CDF Collaboration, F. Abe {\it et al}.,
Phys. Rev. {\bf D50} (1994) 2966; 
Phys. Rev. Lett. {\bf 74} (1995) 2626; 
D0 Collaboration, S. Abachi {\it et al}., 
Phys. Rev. Lett. {\bf 74} (1995) 2632.
%
%
\bibitem{CLEO}
CLEO Collaboration, M.S. Alam, {\it et al}., 
Phys. Rev. Lett. {\bf 74} (1995) 618.
%
\bibitem{Neutralino_Bound}
L3 Collaboration, M. Acciarri {\it et al}., 
Phys. Lett. {\bf B350} (1995) 109.
%
%
\bibitem{LEP2}
D. Treille, 
Talk presented in 29th International Conference on High Energy 
Physics, Vancouver, July 1998.
%
%
\bibitem{CDF+D0}
CDF Collaboration, F. Abe {\it et al}.,
Phys. Rev. Lett. {\bf 75} (1995) 613; 
{\it ibid.}\ {\bf 69} (1992) 3439; 
D0 Collaboration,  S. Abachi {\it et al}., 
Phys. Rev. Lett. {\bf 75} (1995) 618.
%
%
\bibitem{CCB}
J.-P. Derendinger and C. A. Savoy,
Nucl. Phys. {\bf B237}, 307 (1984). 
%
%
\bibitem{Lattice}
N. Tsutsui (JLQCD Collaboration), 
Talk presented in Lattice 98, Boulder, July 1998. 
%
%
\end{thebibliography}
\end{document}